\documentclass[twocolumn,aps,prl,superscriptaddress]{revtex4-2}

\usepackage{orcidlink}  

\usepackage{graphicx}  
\usepackage{bm}        
\usepackage{amssymb}   
\usepackage{xcolor}
\usepackage{amsmath}
\usepackage{natbib}
\usepackage{braket}
\definecolor{linkColor}{RGB}{0,80,150}
\usepackage{hyperref}
\usepackage{algpseudocode}
\usepackage{algorithm}
\usepackage{wrapfig}   

\algrenewcommand\algorithmicrequire{\textbf{Input:}}
\algrenewcommand\algorithmicensure{\textbf{Output:}}
\usepackage{varwidth}

\usepackage[normalem]{ulem}
\usepackage{soul}

\hypersetup{
    colorlinks=true,
    allcolors=linkColor,
    pdfborder={0 0 0},
    pdfencoding = auto
}

\newcommand{\zw}[1]{{\textcolor{black}{ #1 }}}

\makeatletter
\newcommand*{\rom}[1]{\expandafter\@slowromancap\romannumeral #1@}
\makeatother
 
\begin{document}

\title{  Semi-Deterministic Quantum Dot Placement in Heteroepitaxy}

\author{Zihang Wang\,\orcidlink{0000-0003-4482-117X}}

\email{zihangwang@ucsb.edu}

\affiliation{Department of Physics, University of California Santa Barbara, Santa Barbara, California 93106, USA}
 
 \author{Dirk Bouwmeester\,\orcidlink{0000-0002-2118-6532}}

\affiliation{Department of Physics, University of California Santa Barbara, Santa Barbara, California 93106, USA}
 \affiliation{Huygens-Kamerlingh Onnes Laboratory, Leiden University, P.O. Box 9504, 2300 RA Leiden, Netherlands}

\begin{abstract}
Achieving deterministic placement of self-assembled quantum dots (QDs) during epitaxial growth is essential for the reliable and efficient fabrication of high-quality single-photon sources and solid-state cavity quantum electrodynamics (cQED) systems, yet it remains a significant challenge due to the inherent stochasticity of QD nucleation processes. In this work, we theoretically and numerically demonstrate that deterministic QD nucleation within a pristine growth region, e.g., \zw{InAs on a (001)-oriented GaAs} substrate, can be achieved by engineering the boundary geometry of that region. During epitaxial growth, adatoms initially move toward the boundary and promote the formation of primary QDs along the boundary, driven by curvature and diffusion anisotropy. The resulting primary QDs distribution will generate many-body interactions that dynamically reshape the chemical potential landscape for subsequently deposited adatoms, enabling the formation of secondary QDs within the pristine growth region. These findings provide a theoretical foundation for reliable patterning of high optical-quality QDs, with potential applications in next-generation quantum photonic devices.

\end{abstract}

\maketitle

\definecolor{lightblue}{RGB}{200,220,255}
\setlength{\fboxsep}{0.03\linewidth}
\noindent\fcolorbox{white}{white}{\parbox{0.94\linewidth}{%
}}
\vspace{0.5em}

\section{\rom{1}. Introduction }
Self-assembled semiconductor quantum dots (QDs) play a central role in on-chip quantum optics devices. Integrating QDs with optical microcavities, such as photonic crystals~\cite{WOS:000225020200040,Ellis2011,Strauf2006}, microdisks~\cite{WOS:000225020200039}, and micropillars~\cite{WOS:000231017700062,Strauf2007,PhysRevApplied.9.031002,Najer2019}, results in cavity quantum electrodynamics (cQED) systems capable of generating single photons on demand in a single optical mode via the Purcell effect. This capability is essential for implementing quantum cryptographic schemes and quantum communication networks~\cite{WOS:000256839900044,WOS:000457125000008,Morrison2023}.

With additional single electron charge control, QDs can be used to entangle photons with electron spins~\cite{WOS:000311031600042,PhysRevLett.110.037402,PhysRevLett.110.167401,PhysRevLett.104.160503,Kroutvar2004}, enabling coherent transfer of quantum states between spin and photonic degrees of freedom~\cite{PhysRevLett.119.060501,https://doi.org/10.1002/qute.201900085}. Such coherent light matter interactions at the single quantum level are essential for establishing distributed entanglement across quantum networks~\cite{Heindel:23}. Recent achievements include the deterministic on-chip generation of spin photon entanglement in QDs~\cite{WOS:000990997300001}, optical polarization and cooling of nuclear spin ensembles via the optical driving of a charged QD, enabling access and control at the single spin level~\cite{doi:10.1126/science.aaw2906}, and the high efficiency generation of spin photon entangled states, a critical capability for future quantum repeater architectures~\cite{Senellart2017}.

The scaling of this semiconductor technology platform is however hampered by the fact that self-assembled QDs nucleate at random locations on the surface during molecular beam epitaxy (MBE). As a result, microcavities must either be carefully fabricated around pre-characterized QD locations (e.g., Refs.~\cite{Schneider_2009,Sapienza2015}) or post-selection must be used to identify devices in which a QD happens to be well aligned with the cavity mode.

To avoid this time and resource consuming approach, several attempts have been made to implement pre-patterned templates, such as those generated by buried stressors~\cite{Limame2024,Solodovnik2025} and capping layers~\cite{https://doi.org/10.1002/adfm.202304645,Zhao2025}, to control adatom migration so that QD nucleation occurs at predetermined locations. However, the accuracy with which these locations can be predicted is still influenced by surface asymmetric diffusion, strain resulting from lattice mismatch or surface imperfections, and the precise growth parameters~\cite{NISHINAGA2015943,Bart2022,JOYCE1998357,Shchukin1999}. The interplay of these effects remains largely unexplored, especially in efforts to predict QD spatial distributions under realistic growth conditions~\cite{PhysRevB.58.4566,Berdnikov2024}.

Coarse-grained models have captured qualitative features observed in experiments. For example, kinetic Monte Carlo (KMC) simulations have demonstrated how pre-patterned substrates can direct and control nucleation sites~\cite{Urminen2002}; phase-field simulations have reproduced QD densities comparable to experimental results on flat surfaces and have shown that surface patterns can guide QD nucleation, leading to more ordered dot placement~\cite{DelGaudio2018}.

While these theoretical models successfully capture the qualitative modulation of QD density resulting from strain fields, the role of spatiotemporal correlations in QD distributions, such as conditioning on previously formed QDs, has not yet been theoretically addressed.

We propose an overdamped Langevin framework demonstrating that semi-deterministic QD placement is theoretically achievable through the combined effects of geometry-defined boundary profiles and the secondary fields generated by previously nucleated QDs along the boundary. These boundary-anchored QDs dynamically reshape the chemical potential landscape, significantly modifying diffusion kinetics of newly deposited adatoms. This results in secondary QD formation on the pristine region within the boundaries. This stochastic formalism provides a comprehensive model for semi-deterministic spatial organization of QDs, opening new avenues for constructing spatially ordered cQED systems for scalable quantum networks. 

 \begin{figure*}[t!]
    \centering
    \includegraphics{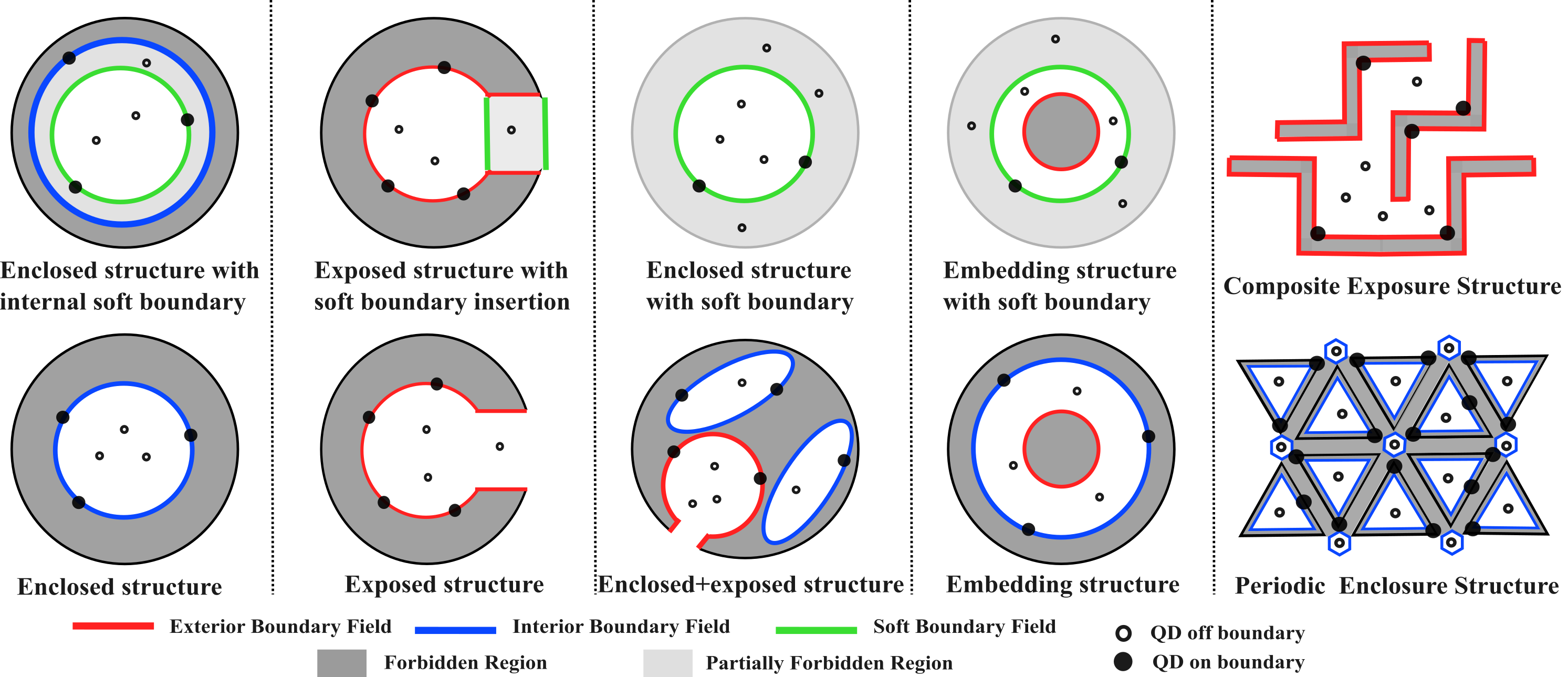}
    \caption{Illustration of patterned geometries used to generate engineered boundary fields (e.g., strain fields). Red and blue lines indicate the boundary contours of patterned regions that induce localized field gradients. For illustrative purposes, solid black circles represent QD nucleating directly at these boundaries, while open circles denote QDs formed under the collective influence of both the boundary fields and QDs already formed on the boundary. Various pattern geometries demonstrate the flexibility of geometric design in controlling the spatial distribution of boundary-induced QD nucleation. The dark gray regions represent areas that adatoms cannot physically access, such as hard masks (forbidden regions), while light gray regions denote areas where adatom access is energetically unfavorable (partially forbidden). We refer to the exposed pristine surface (white regions) as the bulk in this work. } 
    \label{fig1}
\end{figure*}

\section{\rom{2}. Adatom density and QD Nucleation in the Quasi-Stationary Limit }

In traditional heteroepitaxy, atoms are deposited onto a planar substrate with effectively boundaries at infinity, under homogeneous temperature and stress conditions. The spatial adatom density is typically approximated as translationally invariant. As a result, nucleation sites are uniformly distributed across the entire substrate. QD nucleation typically occurs via the Stranski–Krastanov (SK) growth mode \cite{baskaran2012mechanisms, sautter2020strain}. This mode is one of the three primary epitaxial growth modes observed in thin-film deposition and is particularly relevant in heteroepitaxy where two different materials are involved (e.g., InAs on GaAs) \cite{venables1984nucleation}. It describes a transition from two-dimensional (2D) layer-by-layer growth to three-dimensional (3D) island formation. When deposition begins, the adatoms initially form a thin, flat wetting layer on the substrate surface. This layer grows coherently, adopting the lattice constant of the underlying substrate despite a lattice mismatch. As the wetting layer thickens, strain energy accumulates due to lattice mismatch between the film and the substrate (e.g., InAs has a $\sim 7\%$ larger lattice constant than GaAs) \cite{jenkins2014orientation}. \zw{After reaching a critical thickness, typically a few monolayers, local instabilities are induced to release strain energy elastically, triggering spontaneous formation of symmetric three-dimensional islands \cite{tersoff1993shape}. Adatoms begin to aggregate at these energetically favorable sites, forming nanometer-sized clusters (QDs) that are stable (self-limited in size) and exhibit various morphologies and transitions, such as the pyramid-to-dome transition, which have been observed and investigated both experimentally and theoretically \cite{costantini2005pyramids,Stangl2004,long2001effect}. As more adatoms are introduced into the system, coherent (elastic) QD growth becomes metastable, and incoherent (plastic) growth mechanisms, such as dislocation formation, ripening, and the adoption of elongated island shapes \cite{daruka1997dislocation,tersoff1993shape}, emerge. In this work, we focus on the coherent QD growth regime within the pristine surface where QDs are considered dislocation-free and adopt symmetric shapes on the wetting layer.} 

\zw{While the proposed theoretical framework is in general applicable to any substrates with arbitrary orientations, specifically, we address specifically the standard SK growth mode of InAs on the GaAs(001)-$\beta_2 (2\times4)$ surface \cite{rosini2009indium}}: arsenic (As) adatoms tend to form surface dimers upon attachment, resulting in dimer row or valley structures. Indium (In) atoms diffuse with significantly different rates depending on whether their motion is tangential, which is slower and aligned along $\mathbf{\hat{y}} \equiv [1{1}0]$, or orthogonal, which is faster and aligned along $\mathbf{\hat{x}}\equiv  [1\bar{1}0]$, to the As dimer rows~\cite{PhysRevLett.79.5278}. \zw{This biaxial anisotropy is governed by a rank-2 diffusion tensor, which results in elongated quantum dots aligned along the fast diffusion axis $[1\bar{1}0]$~\cite{10.1063/1.2354007}. In this work, we assume that the diffusion tensor is dominated by its diagonal components, although, in general, non-zero off-diagonal terms can contribute to surface diffusion, depending on the substrate crystallographic orientation, adatom species, and temperature.}

As suggested by several recent experimental studies, the introduction of boundary fields may arise from local strain gradients, such as those generated by buried stressors \cite{Limame2024,Solodovnik2025} and capping layers \cite{https://doi.org/10.1002/adfm.202304645,Zhao2025}, or from thermal gradients across the substrate. As illustrated in Fig.\ref{fig1}, we present several examples of boundary fields that can be parametrized either by arc length or by the shape of composite geometries. We classify boundary fields into three types: exterior, interior, and soft boundary fields. Both exterior and interior fields correspond to hard physical boundaries (for example, hard mask edges) that adatoms cannot physically penetrate during the epitaxy process. In contrast, a soft boundary field, such as one induced by buried stressors, is associated with partially forbidden regions where adatom migration is energetically unfavorable during growth, for example, in regions exhibiting thermal or strain gradients. Boundary field strengths, governed by in-plane compressive strain and local impurities, can alter the diffusion dynamics of adatoms during epitaxy, resulting in an increased likelihood of quantum dot nucleation within or near the defined boundary region. We refer to the exposed pristine surface (white-colored regions in Fig.\ref{fig1}) as the bulk.

The dynamics of the adatom density field are governed by the two-dimensional continuity equation, which relates the temporal change in density to the divergence of the adatom flux, 
\begin{equation}
    \partial_t n(\mathbf{r}, t) + \nabla \cdot \mathbf{J}(\mathbf{r}, n, B, t)= 0,
\end{equation}
where $\mathbf{J}(\mathbf{r}, n, B, t)$ is the local adatom current density, $\mathbf{r}$ is the spatial coordinate, $t$ is time, and $B=\int du\,B[\mathbf{r} - \mathbf{r}_{\mathrm{bd}}(u)]$ represents the spatially varying boundary field strength at $\mathbf{r}$. The boundary field is parameterized by $u$, the arc-length variable along the boundary, and $\mathbf{r}_{\mathrm{bd}}(u)$ denotes the unit-speed spatial parametrization of the boundary at $u$. We note that, in general, the boundary field is a tensor that captures structural coupling within the boundary contour. Without loss of generality of our method, we assume that $ B[\mathbf{r} - \mathbf{r}_{\mathrm{bd}}(u)] $ is a scalar field.

\zw{The current density (flux) $\mathbf{J}(\mathbf{r}, n, B, t)$ can be decomposed into two contributions: the lateral current density $\mathbf{J}_{\mathrm{lateral}}$ and the vertical current density $\mathbf{J}_{\mathrm{vertical}}$, representing in-plane adatom diffusion and out-of-plane deposition or desorption, respectively. In the quasi-stationary limit, the epitaxy rate at a given site $\mathbf{r}$ is much smaller than the lateral diffusion rate, such that,
\begin{equation}
|\mathbf{J}_{\mathrm{vertical}}| \sim \lambda n+ J_{\mathrm{dep}} \left(\sum_{m=-\infty}^{\infty} \delta(t - m\, t_{\mathrm{dep}})\right) \ll |\mathbf{J}_{\mathrm{lateral}}|,
\end{equation}
where $\lambda$ is the constant desorption rate, $t_{\mathrm{dep}}$ is the period between subsequent depositions, and $J_{\mathrm{dep}}$ is the deposition flux magnitude. This leads to a separation of fast and slow timescales: the lateral adatom density equilibrates rapidly, while the source and sink terms in the continuity equation associated with deposition or desorption act only as slow-varying perturbations. }

\zw{In the quasi-stationary limit, where $\mathbf{J}(\mathbf{r}, n, B, t)\approx \mathbf{J}_{\mathrm{lateral}}$, the lateral evolution of the adatom density is governed by the coupled dynamics of the local adatom density $n(\mathbf{r})$ and the associated lateral current density $\mathbf{J}$, which can be described by a quasi-equilibrium Boltzmann distribution,}
\begin{equation}
P[n, \mathbf{J}] = \frac{1}{\mathcal{Z}} \exp\!\Big[-\frac{1}{k_{B}T}\, H[n, \mathbf{J}, B]\Big],
\end{equation}
where $H[n(\mathbf{r}), \mathbf{J}(\mathbf{r}), B]$ is a general many-body correlation functional associated with the configuration $[n, \mathbf{J}]$, and $\mathcal{Z}$ is the partition function $\mathcal{Z}=\int \mathcal{D}n   \mathcal{D}\mathbf{J} \exp [- H /{k_B T} ]$, integrated over all admissible field configurations.

The QD nucleation probability at a site $ \mathbf{r} $ depends on the ensemble average over all density and current fields and takes the form,
\begin{equation}
P_{\mathrm{QD}}(\mathbf{r}) = \frac{1}{\mathcal{Z}} \int \mathcal{D}n\, \mathcal{D}\mathbf{J}\; P[n, \mathbf{J}]\,\cdot \nu\big(n(\mathbf{r}), \mathbf{J}(\mathbf{r})\big),
\end{equation}
where $ \nu(n(\mathbf{r}), \mathbf{J}(\mathbf{r})) $ is the local nucleation rate function, modeled as,
\begin{equation}
\nu(n, \mathbf{J}) = \nu_0 \cdot \Theta\big(n - n_{\mathrm{seed}}\big) \cdot \mathcal{R}(n, \mathbf{J}),
\end{equation}
with $ {n}_{\mathrm{seed}} $ denoting the lowest adatom density (subcritical density) that initiates QD nucleation process at a given site. $ \Theta $ is the Heaviside step function, and $ \mathcal{R}(n, \mathbf{J}) $ is a modulation factor that may be modeled as a delta function $ \delta(n - {n}_{\mathrm{seed}}) $, or a smooth exponential dependence on $ n $ and $ \mathbf{J} $.

While this expression for the nucleation probability is formally exact in the quasi-stationary limit, the high-dimensional structure of the functional renders analytical treatment infeasible in most practical scenarios.

Direct sampling of the distribution $ P[n, \mathbf{J}] $ is computationally intractable due to nonlocal correlations and nonlinear coupling between the adatom density $ n(\mathbf{r}) $ and flux $ \mathbf{J}(\mathbf{r}) $ embedded in the functional $ H[n, \mathbf{J}, B] $. Consequently, conventional importance-sampling methods are often inefficient or unstable. To overcome this challenge, we adopt a stochastic sampling approach based on Langevin dynamics, which enables Monte Carlo sampling of field configurations according to the Boltzmann weight $ \exp\left[-H[n, \mathbf{J}, B]/(k_B T)\right] $. In this framework, the time evolution of $ n(\mathbf{r}, t) $ and $ \mathbf{J}(\mathbf{r}, t) $ is governed by coupled Langevin equations that incorporate both deterministic drift toward the minimum of $ H[n, \mathbf{J}, B] $ and stochastic noise consistent with thermal fluctuations, in accordance with the fluctuation–dissipation theorem \cite{PhysRev.83.34}. The steady-state solution of the corresponding Fokker–Planck equation yields the quasi-stationary distribution $ P[n, \mathbf{J}] $ \cite{Risken1989}. This approach allows efficient exploration of the high-dimensional field space and accurate estimation of ensemble-averaged observables, including the spatially resolved QD nucleation probability $ P_{\mathrm{QD}}(\mathbf{r}) $.

\section{\rom{3}. Empirical Field, QD Nucleation, and Langevin-Based Monte Carlo Sampling }
To sample the quasi-stationary field configurations from $ P[n, \mathbf{J}] $, we introduce a discrete empirical adatom density field $ \hat{n}(\mathbf{r}, t) $ and empirical current density $ \hat{\mathbf{J}}(\mathbf{r}, t) $, defined as sums over Dirac delta functions centered on an ensemble of adatoms $ \{\mathbf{r}_i\} $, where each adatom is labeled by an index $ i $,
\begin{equation}
\hat{n}(\mathbf{r}, t) = \sum_{i=1}^{N} \delta\big(\mathbf{r} - \mathbf{r}_i(t)\big), \quad 
\hat{\mathbf{J}}(\mathbf{r}, t) = \sum_{i=1}^{N} \frac{d \mathbf{r}_i(t)}{dt} \, \delta\big(\mathbf{r} - \mathbf{r}_i(t)\big),
\end{equation}
which satisfy the continuity equation by construction. We emphasize that ``empirical” here denotes a representation of the exact microscopic configuration, constructed without any coarse-graining.

 \begin{figure*}[t!]
    \centering
    \includegraphics{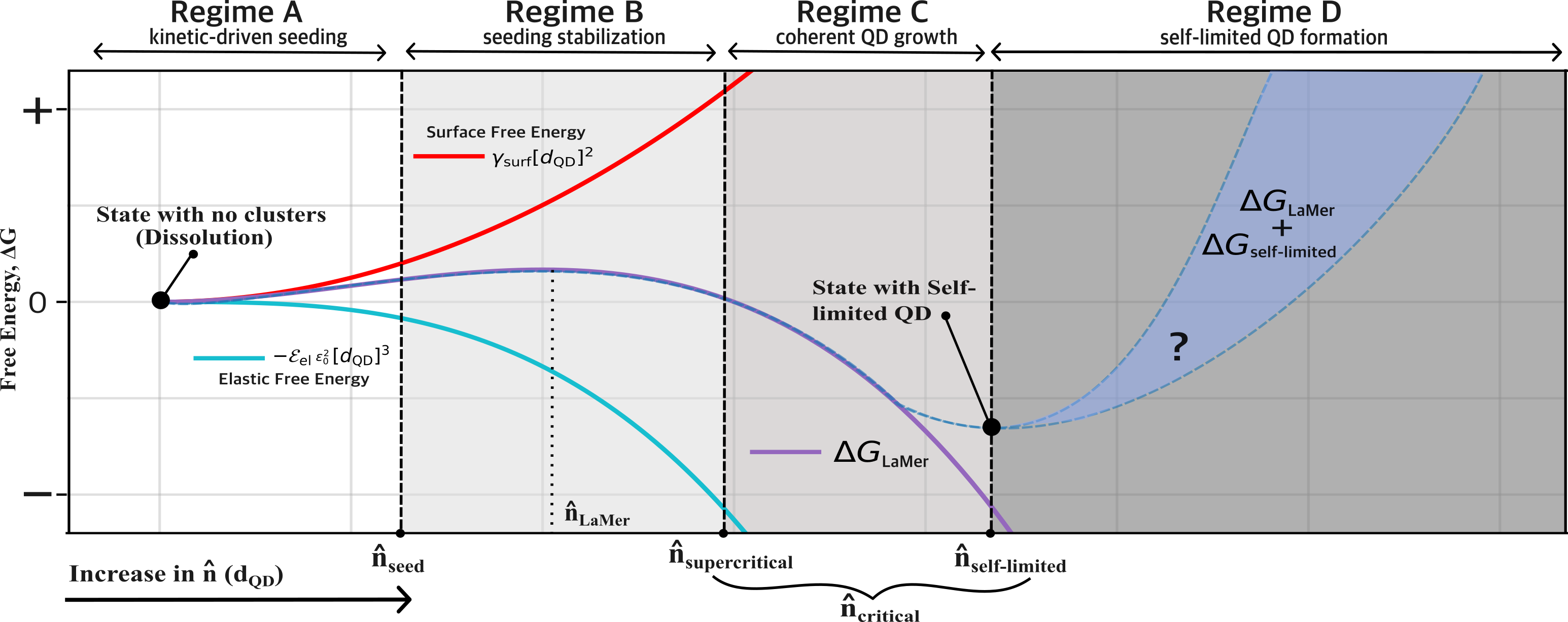}
    \caption{\zw{The free energy landscapes with four regimes (shaded with different colors) of quantum dot formation, detailed in the main text. The formation self-limited QDs exhibits two local minima representing a dissolved state and self-limited QD state, separated by a free energy barrier, peaked (local maximum) at the density $\hat{n}_\mathrm{LaMer}$. As discussed in Eq.~\ref{LaMer eq}, the LaMer free energy $\Delta G_\mathrm{LaMer}$ (purple curve) represents, as a function of cluster density $\hat{n}$, the balance between the free energy contributions from interfacial elasticity (cyan curve) and the formation of the cluster surface (red curve). Four regimes (A-D) are separated by three characteristic densities, subcritical density $\hat{n}_{\mathrm{seed}}$, supercritical density $\hat{n}_{\mathrm{supercritical}}$, and self-limited density $\hat{n}_{\mathrm{self-limited}}$, associated with the formation of temporary clusters, clusters lead to coherent QD growth, and self-limited (stable) QDs. In regime D (shaded in blue), the net free energy $\Delta G$ given in Eq.~\ref{total E eq} may exhibit a convex behavior around the self-limited (stable) QDs. $\hat{n}_{\mathrm{supercritical}}$ and $\hat{n}_{\mathrm{self-limited}}$ are referred as a single critical density $\hat{n}_{\mathrm{crit}}$.}} 
    \label{fig2}
\end{figure*}

We assume that adatoms are initially deposited uniformly onto the surface $ \Omega $, so that $ \hat{n}(\mathbf{r}, 0) \sim 1/{\Omega} $, with ${\Omega}$ being the surface area, and each adatom $ i $ undergoes stochastic evolution governed by the following overdamped Langevin dynamics $(k_B T\equiv1)$,
\begin{equation}
m \frac{d^2 \mathbf{r}_i}{dt^2} \approx 0
= - \boldsymbol{\Gamma} \, \frac{d\mathbf{r}_i }{dt}
- \nabla_i U_{\mathrm{sys}}
+ \sqrt{2 \, \boldsymbol{\Gamma}} \; \boldsymbol{\eta}_i(t),
\label{single L}
\end{equation}
where $ m $ is the effective mass, $ \boldsymbol{\Gamma} $ is the friction tensor (assumed diagonal) along the two principal crystallographic directions $[1\bar{1}0]$ and $[110]$, and $ \boldsymbol{\eta}_i(t) $ is a Gaussian noise term (a \textit{distribution}) with statistics $ \langle \eta_i(t)\, \eta_j(t') \rangle = \delta_{ij} \, \delta(t - t') $.

The total microscopic potential energy $ U_{\mathrm{sys}} $ includes both the boundary field $ B $ and repulsive pairwise adatom interaction $ w $, defined as a functional of the empirical adatom density field over the entire exposed substrate area $ \Omega $, 
\begin{equation}
\begin{aligned}
U_{\mathrm{sys}}[\hat{n}] &= \int du \int_{\Omega} B[\mathbf{r} - \mathbf{r}_{\mathrm{bd}}(u)] \, \hat{n}(\mathbf{r}, t) \, d\mathbf{r} \\
&\quad + \frac{1}{2} \iint_{\Omega} \hat{n}(\mathbf{r}, t) \, w(\mathbf{r} - \mathbf{r}') \, \hat{n}(\mathbf{r}', t) \, d\mathbf{r} \, d\mathbf{r}'.
\end{aligned}
\label{full energy}
\end{equation}

By summing over all adatoms $ i $, the empirical density field evolves according to the following stochastic differential equation (SDE),
\begin{equation}
\begin{aligned}
    \frac{\partial \hat{n}(\mathbf{r}, t)}{\partial t}
=& \nabla \cdot \left[ \hat{n}(\mathbf{r}, t) \mathbf{D}  \nabla \mu(\mathbf{r}, t) \right]+ \nabla \cdot \left[ \sqrt{2  \hat{n}(\mathbf{r}, t) \mathbf{D} } \, \boldsymbol{\eta}(\mathbf{r}, t) \right],
\end{aligned}
\label{SDE for n}
\end{equation}
where the diffusion tensor is given by $ \mathbf{D} ={   \boldsymbol{\Gamma}^{-1}} $, and \zw{the functional derivative $ \mu = \delta U_{\mathrm{sys}} / \delta \hat{n}(\mathbf{r}, t) $ represents the single-adatom chemical potential landscape, as the sum of a boundary-induced term and pairwise interaction contributions,
\begin{equation}
\begin{aligned}
 \mu(\mathbf{r}, t) = &\int du \, B\big[\mathbf{r} - \mathbf{r}_{\mathrm{bd}}(u)\big] 
+ \int_{\Omega} w\big(\mathbf{r} - \mathbf{r}'\big) \, \hat{n}(\mathbf{r}', t) \, d\mathbf{r}'.
\end{aligned}
\end{equation}
The spatial gradient of the above chemical potential landscape $ \mathbf{F}(\mathbf{r}, t)=-\nabla \mu(\mathbf{r}, t) $ corresponds to the microscopic force acting on the adatom ensemble.} While the functional derivative $ \mu = \delta U_{\mathrm{sys}} / \delta \hat{n}(\mathbf{r}, t) $  is not rigorously defined due to the singular nature of $ \hat{n} $, which consists of Dirac delta functions, in practice, the values and gradients of $\mu(\mathbf{r}, t)$ are computed directly from the adatom-level Langevin dynamics. Thus, Eq.~\ref{SDE for n} should be interpreted as describing the stochastic evolution of an adatom ensemble, rather than as a coarse-grained partial differential equation (PDE).  

To sample the density field $ \hat{n}(\mathbf{r}, t) $, we do not solve Eq.~\ref{SDE for n} directly. Instead, we simulate Langevin dynamics for an ensemble of $ N $ adatoms and estimate the empirical field via a grid-based Monte Carlo technique. Specifically, we evolve the single-adatom Langevin equation given in Eq.~\ref{single L},
\begin{equation}
\frac{d\mathbf{r}_i}{dt} = 
- \mathbf{D} \nabla_i U_{\mathrm{sys}} 
+ \sqrt{2  \, \mathbf{D}} \; W_i(t),
\label{adatom stochastic}
\end{equation}
where $ W_i(t) $ is a standard Wiener process. From the resulting trajectories, the empirical adatom density field is computed by counting particles on a spatial grid element $ \mathrm{d}\Omega $, parametrized by the two-dimensional variable $ \boldsymbol{l} $, centered at $ \mathbf{r}_0 $ and covering the exposed substrate region $ \Omega $. This yields a time-dependent realization of the empirical field that satisfies the above SDE over ensemble averages,
\begin{equation}
\hat{n}(\mathbf{r}) \approx \frac{1}{\mathrm{d}\Omega} \sum_{i} \int_{\mathrm{d}\Omega} \mathrm{d}\boldsymbol{l} \, \delta(\mathbf{r}_i - \mathbf{r} + \boldsymbol{l}).
\end{equation}

The above Langevin formulation, which generates the empirical adatom density field, must also incorporate microscopic nucleation processes that lead to QD formation during heteroepitaxy. To account for this, we define the local nucleation free energy $ G(\hat{n}, \mathbf{r}_0) $ as the free energy cost to assemble a cluster of $ \hat{n} $ adatoms within a grid cell of width $ \sqrt{d\Omega}$, centered at $ \mathbf{r}_0 $. This nucleation barrier is computed by evaluating the system energy $ U_{\mathrm{sys}}[\hat{n}(\mathbf{r}, t)] $ on a constrained configuration in which $ \hat{n} $ adatoms are localized near $ \mathbf{r}_0 $, and subtracting the reference energy of the homogeneous background state $ \hat{n}(\mathbf{r}, 0) \sim 1/{\Omega} $,
\begin{equation}
G (\hat{n}, \mathbf{r}_0) = U_{\mathrm{sys}}[\hat{n}(\mathbf{r}_0, t)] - U_{\mathrm{sys}}[\hat{n}(\mathbf{r}_0, 0)].
\end{equation}
The function $ G(\hat{n}, \mathbf{r}_0) $ captures the energetic trade-off between inter-adatom interactions, strain relaxation, and the influence of boundary fields near $ \mathbf{r}_0 $. \zw{We denote the energy $U_{\mathrm{sys}}[\hat{n}(\mathbf{r}_0, t)]$ as the free energy, as the entropic contribution (compared to the strain and surface energies) is negligible under typical growth conditions \cite{daruka1997dislocation}.}

\zw{The apparent complexity of the above free energy functional can be simplified within the coherent SK growth mode by adopting a classical LaMer nucleation model \cite{D0MA00439A,tersoff1994competing}, which captures the essential competing mechanisms during coherent QD growth, and QDs are considered dislocation-free and exhibit symmetric shapes. Since the QD volume is directly proportional to the number of adatoms $\hat{n}$, for QD formation at a site $\mathbf{r}_0$ with an approximate nucleus diameter $d_{\mathrm{QD}} \sim \hat{n}^{1/3}$, the following model describes the competition between elastic strain energy relief and surface free energy, with magnitudes $\mathcal{E}_{\mathrm{el}}$ and $\gamma_{\mathrm{surf}}$, respectively,
\begin{equation}
    \Delta G_{\mathrm{LaMer}}(d_{\mathrm{QD}})=-\epsilon_0^2\mathcal{E}_{\mathrm{el}}[d_{\mathrm{QD}}]^{3}+ \gamma_{\mathrm{surf}} [d_{\mathrm{QD}}]^{2},
    \label{LaMer eq}
\end{equation}
where $\epsilon_0=(a_{\mathrm{film}}-a_{\mathrm{sub}})/a_{\mathrm{sub}}$ is the lattice mismatch between the film ($a_{\mathrm{film}}$) and the substrate ($a_{\mathrm{sub}})$ lattice constants. The first term corresponds to the elastic strain energy relief with respect to the wetting layer, and the second term captures the energetic penalty of generating QD surfaces \cite{tersoff1994competing}. Motivated by the experimental observation of self-limited QDs \cite{Stangl2004}, we introduce a phenomenological free energy term $\Delta G_{\mathrm{self-limited}} > 0$, which limits coherent QD growth and accounts for the emergence of a second local minimum in the free energy landscape,
\begin{equation}
\begin{aligned}
     \Delta G(d_{\mathrm{QD}}, \mathbf{r}_0) & \sim \Delta G_{\mathrm{LaMer}}(d_{\mathrm{QD}}, \mathbf{r}_0)+\Delta G_{\mathrm{self-limited}}(d_{\mathrm{QD}}, \mathbf{r}_0).
\end{aligned}
\label{total E eq}
\end{equation}
While attempts have been made in obtaining the approximated form of this term \cite{li2008thermodynamic}, the exact mechanisms underlying self-limited QD sizes remain under debate (e.g., the competition between kinetic and thermodynamic processes) and are beyond the scope of this work \cite{wu2015self,barabasi1999thermodynamic}. Nevertheless, regardless of the detailed form of this free energy term, its existence, along with the engineered boundary field, plays a crucial role in manipulating the empirical density field of subsequently deposited adatoms, enabling semi-deterministic QD placement on the bulk surface.}

\zw{As illustrated in Fig. \ref{fig2}, we present four regimes of the free energy $\Delta G$ that capture essential aspects of QD formation: 
\begin{itemize}
    \item Regime A (kinetic-driven seeding): Adatoms are continuously deposited onto the bulk surface, where they stochastically migrate and contribute to the formation of surface layers on top of the substrate. Upon approaching the critical thickness, adatoms form temporary seeding clusters with subcritical density $\hat{n}_{\mathrm{seed}}$ as the starting seed of QD growth. 
    \item Regime B (seeding stabilization): Upon reaching the subcritical bulk volume, further adatom deposition leads to the formation of transient seeding clusters that are energetically favorable. These clusters may increase in local density $\hat{n}$ as they grow elastically. However, they remain susceptible to dissolution through thermal fluctuations or Ostwald ripening, in which smaller clusters are absorbed by larger ones to reduce the system’s total surface energy, until they eventually exceed a supercritical density $\hat{n}_{\mathrm{supercritical}}$  \cite{wu2015self}. 
    \item Regime C (coherent QD growth): Once clusters exceed the supercritical size, they grow continuously until reaching a self-limited density $\hat{n}_{\mathrm{self-limited}}$, at which point they become stable QDs. We note that the lattice mismatch must be sufficiently large to support stable QDs \cite{daruka1997dislocation}. Although the stabilization mechanisms are still in debate, the second minimum of the free energy, $\Delta G_{\mathrm{LaMer}} + \Delta G_{\mathrm{self-limited}}$, accounts for the existence of self-limited QDs. In contrast, the classical LaMer nucleation model, $\Delta G_{\mathrm{LaMer}}$, predicts unbounded QD growth corresponding to a global minimum. Nevertheless, both the existence of self-limited QDs and their influence on adatoms and neighboring QDs have been investigated extensively, both theoretically and experimentally. 
    \item Regime D (self-limited QD formation): QDs reach a stationary size, and the addition of further adatoms is no longer energetically favorable (i.e., the QDs become metastable toward plastic growth), generating an effective repulsive force on nearby adatoms \cite{barabasi1999thermodynamic}. These QDs act as sources of strain fields that elastically distort the wetting layer and substrate \cite{koduvely1999epitaxial}. As a consequence, a uniform spatial distribution of QDs emerges, as observed both experimentally and theoretically via kinetic Monte Carlo simulations \cite{kobayashi1996in,meixner2003control}.
\end{itemize}}

\zw{While obtaining all microscopic parameters associated with the above four regimes is challenging, we can obtain valuable insights on how to manipulate QDs by considering the following simplifications together with the aforementioned Langevin dynamics, Eq.~\ref{adatom stochastic}. The above four regimes can be formalized through a time-dependent nucleation potential centered at $ \mathbf{r}_0 $, evolving as,
\begin{equation}
V_{\mathrm{nc}} (\mathbf{r} - \mathbf{r}_0, t) = 
\begin{cases}
V_{\mathrm{seed}}(\mathbf{r} - \mathbf{r}_0, t), 
& \text{if } |t - t_0| \leq \tau, \\[1em]
V_{\mathrm{QD}}(\mathbf{r} - \mathbf{r}_0), 
& \text{if } \hat{n}(\mathbf{r}_0, t) \geq \hat{n}_{\mathrm{self-limited}}, \\[1em]
0, 
& \text{otherwise (dissolution)},
\end{cases}
\label{nc}
\end{equation}
To model regimes A and B, we introduce a localized seed potential $ V_{\mathrm{seed}} $ at $ \mathbf{r}_0 $ whenever the empirical density exceeds a subcritical threshold $ \hat{n}_{\mathrm{seed}} $. The seeding potential serves as a binding potential $V_{\mathrm{seed}}$ that anchors nearby adatoms, promoting local adatom accumulation and cluster formation. }

\zw{We model $V_{\mathrm{seed}}$ as a radially symmetric trapping potential with sufficient depth (relative to thermal fluctuations), and the effective area of this potential is taken to be the underlying Monte Carlo grid size $d\Omega$. This seed persists for a finite duration $ \tau_c $, during which the local seed is allowed to either evolve toward the QD self-limited size with density $ \hat{n}_{\mathrm{self-limited}} $ at $ \mathbf{r}_0 $, corresponding to the second minimum in $\Delta G$, or dissolve back to the uniform state (no clusters). The seed lifetime $ \tau $ is drawn from an exponential distribution, $ \tau \sim \exp (1/\tau_c) $, and $ t_0 $ denotes the first-passage time for seeding, defined as,
\begin{equation}
t_0 = \min \left\{ t \;\middle|\; \hat{n}(\mathbf{r}_0, t) \geq \hat{n}_{\mathrm{seed}} \right\}.
\end{equation}
If no QD nucleation occurs within the interval $ [t_0, t_0 + \tau] $, the seed potential is removed, corresponding to dissolution. Both the subcritical threshold $ \hat{n}_{\mathrm{seed}} $ and supercritical density $ \hat{n}_{\mathrm{supercritical}} $ determined by detailed growth conditions. Since reaching the supercritical density $\hat{n}_{\mathrm{supercritical}}$ triggers coherent QD growth toward the self-limited density $\hat{n}_{\mathrm{self-limited}}$, for computational simplicity we merge these two points into a single reference and denote it as the critical density $\hat{n}_{\mathrm{crit}}$ (see Fig.~\ref{fig2}). The supercritical and critical densities are taken to be of the same order of magnitude as the homogeneous background state, $\hat{n}_{\mathrm{crit}} \approx 10^3\hat{n}_{\mathrm{seed}}$ and $\hat{n}_{\mathrm{seed}}\approx 10^2\hat{n}(\mathbf{r}_0 , 0)$. }

\zw{As the empirical density reaches $\hat{n}_{\mathrm{crit}}$ within the time window $\tau_c$, the seed potential is irreversibly converted into a stable QD potential $V_{\mathrm{QD}}$, indicating successful traversal of the nucleation barrier into regime C, creating a permanent QD at $\mathbf{r}_0 $ where it subsequently generate repulsion ($V_{\mathrm{QD}}$) to newly deposited adatoms, reflecting the stabilization of the dot and the inhibition of further growth beyond its self-limited size.  }

\zw{The detailed interatomic interactions among adatoms, between adatoms and QDs, and among QDs depend on the growth conditions and specific atomic properties, such as bond ionicity and van der Waals interactions, and are typically obtained from empirical studies \cite{Stangl2004}. For the purpose of demonstrating semi-deterministic QD placement, we model the pairwise repulsion between adatoms (dipole–dipole) and between adatoms and QDs (dipole–monopole) \cite{shilkrot1997adatom}, denoted $V_{\mathrm{nc}}$, using a generic screened Coulomb-like potential $V$, expressed as,
\begin{equation}
V\big(\mathbf{r} - \mathbf{r}'\big) \sim \frac{\exp\left[-\xi\left|\boldsymbol{\ell}^{-1} (\mathbf{r} - \mathbf{r}')\right|\right]}{\left|\boldsymbol{\ell}^{-1} (\mathbf{r} - \mathbf{r}')\right|^{\alpha}}.
\label{pair eq}
\end{equation}
Here, $\xi$ is a screening parameter that sets the scale, and the characteristic decay length-scale tensor is defined as $\boldsymbol{\ell} = \sqrt{\mathbf{D}/\lambda}$, where $\lambda$ is the constant desorption rate that sets the interaction scale. The power-law exponent $\alpha$ is a microscopic parameter that characterizes the type and range of the many-body interactions. For example, a dipole–monopole interaction corresponds to $\alpha = 2$, and a dipole–dipole interaction corresponds to $\alpha = 3$.}

\zw{The system energy is updated by incorporating the time-dependent nucleation potential,
\begin{equation}
U_{\mathrm{sys}} \to U_{\mathrm{sys}} + \frac{1}{2} \iint_{\Omega} \hat{n}(\mathbf{r}, t) \, V_{\mathrm{nc}}(\mathbf{r} - \mathbf{r}_0, t) \, \hat{n}(\mathbf{r}_0, t) \, d\mathbf{r} \, d\mathbf{r}_0,
\end{equation}
and the Langevin dynamics is modified accordingly. The pairwise interaction $w$ in $U_{\mathrm{sys}}$ (Eq.~\ref{full energy}) is approximated as a point-like elastic dipole interaction at long range and is assumed to be repulsive. This discourages adatoms from clustering too early, favors nucleation at distinct sites, and contributes to the size uniformity of QDs \cite{Stangl2004,shilkrot1997adatom}.}

The boundary field is modeled as a delta-function sink potential, modulated by the (dimensionless) local strain profile $ \sigma[\mathbf{r}_{\mathrm{bd}}(u)] $, which prevents adatoms from escaping once they reach the boundary,
\begin{equation}
B\big[\mathbf{r} - \mathbf{r}_{\mathrm{bd}}(u)\big] =-B_0\,\sigma[\mathbf{r}_{\mathrm{bd}}(u)] \, \delta\big[\mathbf{r} - \mathbf{r}_{\mathrm{bd}}(u)\big].
\end{equation}
This formulation allows adatoms to migrate tangentially along the boundary. The sign of the strain profile $ \sigma $ encodes the local stress state: compressive strain (positive $ \sigma $) attracts adatoms, while tensile strain (negative $ \sigma $) repels them. While in general, the boundary field strength would vary due to inhomogeneity along the boundary profile, we assume that the strain results from local lattice mismatch and thus depends only on the local boundary curvature.

\section{\rom{4}. Linear Perturbation Theory in the diluted limit}
\begin{figure*}[t!]
    \centering
    \includegraphics{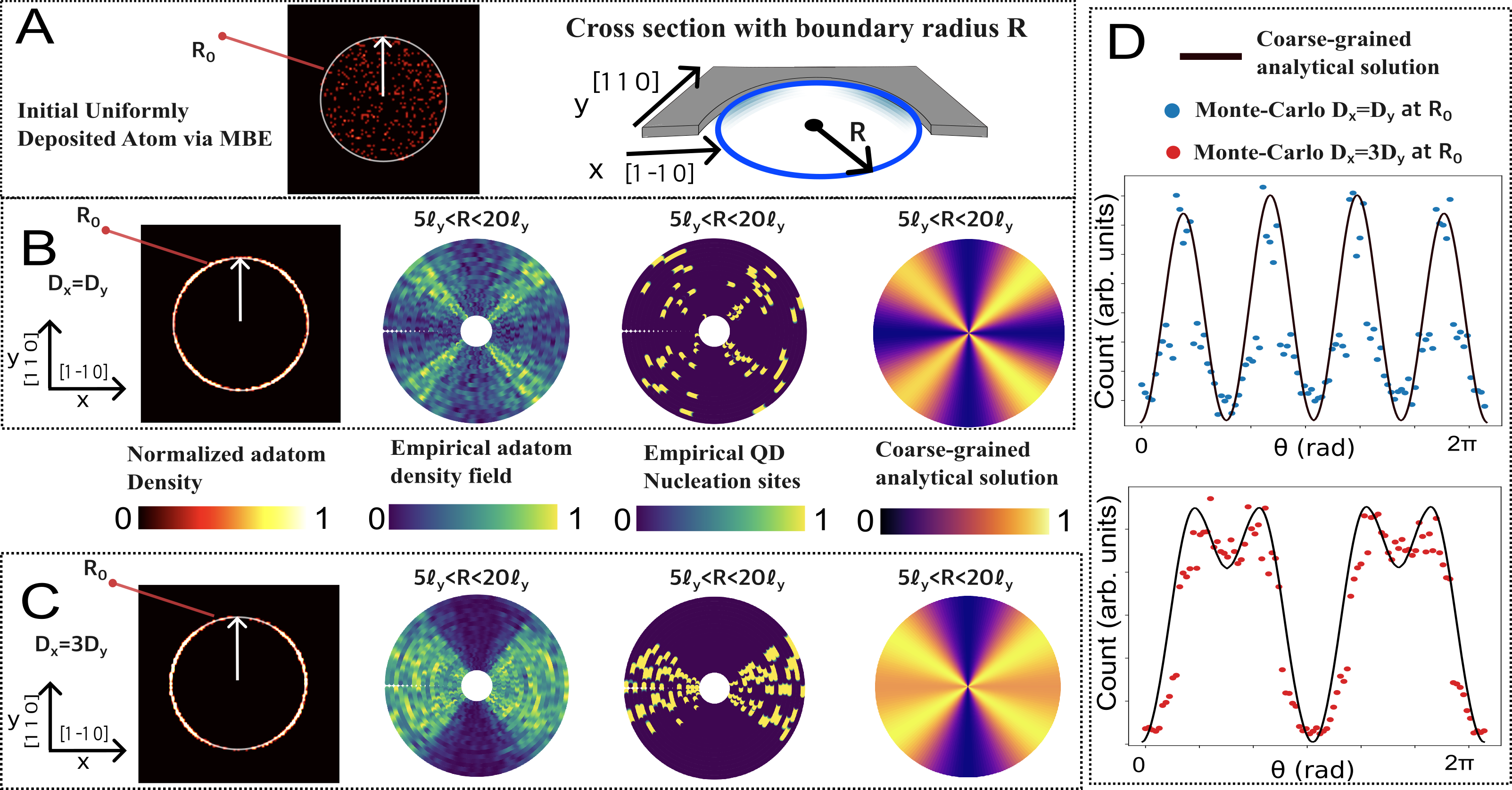}
    \caption{{Detailed comparison of empirical adatom density field (Monte Carlo sampled), QD nucleation, and mean-field analytical adatom density field solutions under a homogeneous circular boundary field, for both isotropic diffusion ($ D_x = D_y $) and anisotropic diffusion ($ D_x = 3D_y $) along the boundary.} \textbf{A}: Atomic flux is uniformly deposited from an epitaxy source onto the center of a circular patterned substrate with radius $ R_0=10\, \ell_y $, {where $\ell_y$ is the diffusion length in $\mathbf{\hat{y}}$ direction}. \textbf{B--C}: The first (leftmost) column shows the empirical adatom density field at a given radius $ R_0=10\, \ell_y $. Due to diffusion, adatoms accumulate at the boundary, and the resulting spatial distribution depends on the diffusion anisotropy. For isotropic diffusion, the empirical density exhibits radial symmetry, while for anisotropic diffusion, the distribution becomes directionally biased. The second column shows radial statistics extracted from various radii with respect to the diffusion length $ 5\, \ell_y <R< 20 \, \ell_y$, further illustrating the breakdown of axial symmetry under anisotropy. In both columns, pairwise interactions are turned off ($ w = 0 $) and the nucleation potential is absent ($ V_{\mathrm{nc}} = 0 $). The third column incorporates a nonzero nucleation potential ($ V_{\mathrm{nc}} \neq 0 $) and displays the resulting QD nucleation sites for various radii $5\, \ell_y <R< 20 \, \ell_y $. Nucleation events are strongly correlated with regions of high adatom density. The fourth column presents mean-field analytical solutions obtained from Eq.~\ref{path-integral}. \textbf{D}: Cross-section comparison between mean-field analytical solutions and Langevin-based Monte Carlo simulations at a given radius $ R_0=10\, \ell_y $, evaluated with respect to angular parametrization under both isotropic and anisotropic conditions.
} 
    \label{fig3}
\end{figure*}

In the early stage of heteroepitaxy, thermally driven stochastic migration of adatoms causes preferential deposition at boundary edges orthogonal to the fast diffusion axis $\mathbf{\hat{x}} \equiv [1\bar{1}0]$ \cite{https://doi.org/10.1002/adfm.202304645}, followed by lateral spreading along the boundary, as shown numerically in Fig.~\ref{fig3}C. When the average empirical density $ \hat{n} $ over the exposed region $ \Omega $ remains much smaller than the homogeneous density limit, the adatom dynamics are primarily governed by anisotropic diffusion. We refer to this regime as the \emph{diluted limit}, where the exposed region $ \Omega $ remains effectively adatom-free.

It is important to note that, locally, the empirical density may exceed this limit on the boundary. Depending on the repulsion strength between adatoms and the critical seeding density, such local accumulation can initiate the early stages of quantum dot nucleation. 

We coarse-grain the empirical field $ \hat{n} $ into a continuous density field $ n $, and introduce a decay term to account for the finite lifetime of adatoms, characterized by a constant desorption rate $ \lambda $. Expanding locally around a boundary point $ \mathbf{r}_{0,\mathrm{bd}} = (x_0, y_0) $, we assume that the convection–reaction–diffusion dynamics are separable to linear order along the principal axes. This yields a local density field $ n_0 $, modulated by separable spatial factors $ f(x - x_0) $ and $ g(y - y_0) $, such that,
\begin{equation}
n(\mathbf{r}_{\mathrm{bd}}) = f(x - x_0) \cdot g(y - y_0) \cdot n_0 
 \Rightarrow  
\ln n = \ln f + \ln g + \ln n_0.
\end{equation}

Near the boundary point $ \mathbf{r}_{0,\mathrm{bd}} $, the change in current density along the boundary curve is given by
\begin{equation}
    \frac{d\mathbf{J}}{ds}(\mathbf{r}_{\mathrm{bd}}) = \frac{dJ_{\parallel}}{ds} \, \mathbf{\hat{T}} + J_{\parallel} \, \frac{d\mathbf{\hat{T}}}{ds} = \frac{dJ_{\parallel}}{ds} \, \mathbf{\hat{T}} + \kappa J_{\parallel} \, \mathbf{\hat{N}},
\end{equation}
where $ \kappa(\mathbf{r}_{0,\mathrm{bd}}) $ is the local curvature, describing how rapidly the boundary turns at $ \mathbf{r}_{0,\mathrm{bd}} $. The vectors $ \mathbf{\hat{T}} $ and $ \mathbf{\hat{N}} $ are the local tangent and outward normal vectors, respectively, evaluated at the boundary.

The tangential component of the current density is defined as,
\begin{equation}
J_{\parallel}(\mathbf{r}_{\mathrm{bd}} - \mathbf{r}_{0,\mathrm{bd}}) = -\mathbf{\hat{T}} \cdot \mathbf{D} \nabla n, \quad \kappa(\mathbf{r}_{0,\mathrm{bd}}) \cdot \mathbf{\hat{T}} \cdot \mathbf{D} \nabla n \geq 0,
\end{equation}
and, for consistency, we adopt the convention that the adatom flow aligns with the direction in which the boundary turns. This convention does not affect the physical direction of adatom motion but serves to interpret the sign of the local current density magnitude. Positive values correspond to accumulation, while negative values indicate depletion, with the sign further modulated by the local strain profile.

The local continuity equation at the boundary point $ \mathbf{r}_{0,\mathrm{bd}} $ is given by,
\begin{equation}
\partial_t n = -\mathbf{\hat{T}} \cdot \nabla J_{\parallel} - \kappa \, \mathrm{Sign}[\sigma(\mathbf{r}_{0,\mathrm{bd}})] \, J_{\parallel} - \lambda n,
\label{equation dym}
\end{equation}
where the sign of the local strain profile, $ \mathrm{Sign}[\sigma(\mathbf{r}_{0,\mathrm{bd}})] $, determines the nature of the curvature-driven contribution: compressive strain ($ \mathrm{Sign} = +1 $) promotes adatom accumulation over time, while tensile strain ($ \mathrm{Sign} = -1 $) drives depletion.

Substituting the expression for $ J_{\parallel} $ into the Eq.~\ref{equation dym}, we have, 
\begin{equation}
\mathbf{\hat{T}} \cdot \nabla \left( \mathbf{\hat{T}} \cdot \mathbf{D} \nabla n \right)
+ \mathrm{Sign}[\sigma(\mathbf{r}_{0,\mathrm{bd}})] \, \kappa \left( \mathbf{\hat{T}} \cdot \mathbf{D} \nabla n \right)
- \lambda n = 0,
\label{pde}
\end{equation}
which can be separated into a system of two PDEs for $ f(x - x_0) $ and $ g(y - y_0) $, respectively.

Expanded in Cartesian components around the boundary point $ \mathbf{r}_{0,\mathrm{bd}} $, the local convection–diffusion–reaction equations take the form,
\begin{equation}
\begin{aligned}
\frac{\partial f}{\partial t} &=
T_x^2 D_x \frac{\partial^2 f}{\partial x^2}
+ \kappa T_x D_x \frac{\partial f}{\partial x}
- \lambda f, \\
\frac{\partial g}{\partial t} &=
T_y^2 D_y \frac{\partial^2 g}{\partial y^2}
+ \kappa T_y D_y \frac{\partial g}{\partial y}
- \lambda g,
\end{aligned}
\label{pde_components}
\end{equation}
where $ T_x = \mathbf{\hat{T}} \cdot \mathbf{\hat{x}} $ and $ T_y = \mathbf{\hat{T}} \cdot \mathbf{\hat{y}} $ denote the projections of the local tangent direction along the two principal diffusion axes. These equations are invariant under parity transformation ($ x \to -x $, $ y \to -y $).

The first-order expansion near a boundary point elucidates the interplay between anisotropic diffusion and geometric modulation, yielding a tractable expression for the stationary adatom density profile along and near the boundary. The general steady-state solution of the adatom density field can be constructed as a product of first-order expansions using the boundary parametrization $u$, where $\theta$ is the corresponding angular variable,
\begin{equation}
\begin{aligned}
\frac{n(u_{N})}{n_0} = 
\prod_{i=0}^{N} \sum_{c, c' = \pm 1} A^i_{c, c'} \exp\left[ du_i \left(k_x^{c} |T_x(u_i)| +  k_y^{c'} |T_y(u_i)| \right)\right],
\end{aligned}
\label{steady-state}
\end{equation}
where $A^i_{c, c'}$ is a coefficient determined by boundary conditions and local geometry, and $du_i$ denotes the local boundary segment at $u_i$. Recall that $\mathbf{r}_{\mathrm{bd}}(u)$ is a unit-speed trajectory ($|d\mathbf{r}_{\mathrm{bd}}(u)/du| = 1$), we have,
\begin{equation}
  \frac{dv_i}{du_i} = |T_v(u_i)| \approx \frac{|v(\theta_{i+1}) - v(\theta_{i})|}{du_i}, \quad v=x, y\,,
\end{equation}
where $v=x, y$ labels the primary diffusion axes. The exponent in Eq.~\ref{steady-state} can be compactly written as (Appendix A),
\begin{equation}
\begin{aligned}
&k_v^{c} |T_v(u_i)| = \frac{\kappa(u_i)}{2} + c\,\mathcal{K}_v(u_i), \\
&\ell_v = \sqrt{ \frac{D_v}{\lambda} },\quad \mathcal{K}_v(u_i) = \sqrt{ \left[ \frac{\kappa(u_i)}{4} \right]^2 + \ell_v^{-2} },
\end{aligned}
\end{equation}
where $D_x$ and $D_y$ are the diagonal elements of the diffusion tensor $\mathbf{D}$, and $\ell_x,\,\ell_y$ define the characteristic decay lengths along the two principal diffusion axes.

In the continuum limit, the general steady-state solution assumes a path-integral form,
\begin{equation}
\begin{aligned}
\frac{n(u)}{n_0} = \int \mathcal{D}[\mathbf{c}, \mathbf{c'}] \exp\Big[ &\int_{0}^{u} d\tilde{u} ( \kappa(\tilde{u})+ \mathbf{c}(\tilde{u})\, \mathcal{K}_x(\tilde{u})\\&  + \mathbf{c'}(\tilde{u})\, \mathcal{K}_y(\tilde{u}) ) \Big],
\end{aligned}
\label{path-integral}
\end{equation}
where the functional integral $\int \mathcal{D}[\mathbf{c}, \mathbf{c'}]$ represents the sum over all possible sign trajectories for $(\mathbf{c}, \mathbf{c'})$,
\begin{equation}
\int \mathcal{D}[\mathbf{c}, \mathbf{c'}]\,\cdots = \iint \mathcal{D}[\mathbf{c}]\, \mathcal{D}[\mathbf{c'}]\, A[\mathbf{c}, \mathbf{c'}]\, \cdots.
\end{equation}

We compare the analytical steady-state adatom density field given in Eq.~\ref{path-integral}, under the mean-field approximation detailed in the Appendix B, for both the biaxial diffusion–dominated condition (i.e., circles with constant curvature) and the curvature-dominated condition (i.e., ellipses with variable curvature). To model the adatom density on the boundary, atoms are initially deposited uniformly within a fixed geometry to mimic the initial epitaxial conditions. The boundary field $ B $ acts as a sink potential: once adatoms reach the boundary, they migrate only along it via a projected Langevin dynamics, in contrast to the full 2D Langevin dynamics that govern motion in the bulk.

 \begin{figure*}[t!]
    \centering
    \includegraphics{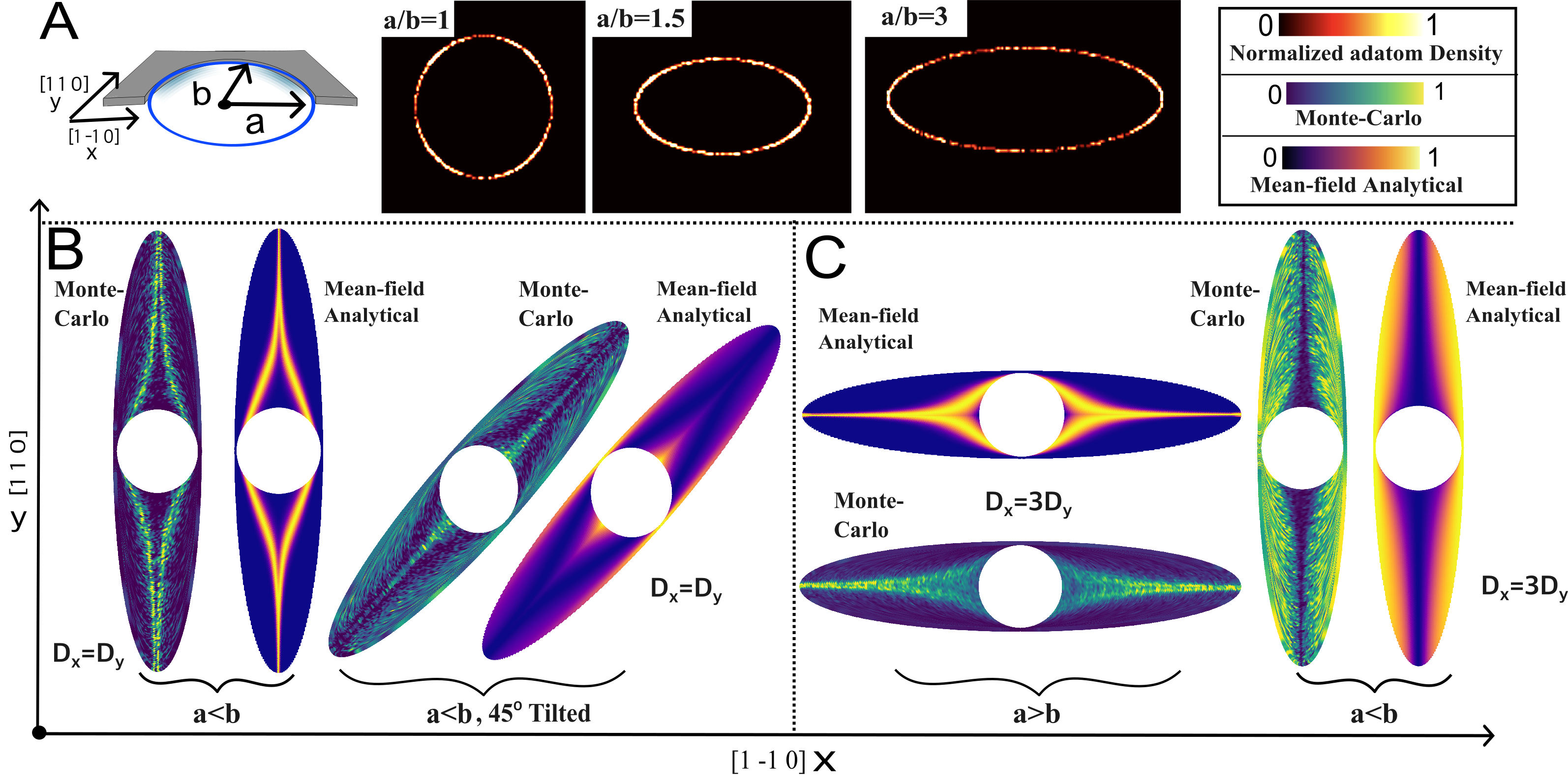}
    \caption{{Detailed comparison between empirical adatom density fields (Monte Carlo) and mean-field analytical adatom density field solutions under a homogeneous elliptical boundary field for isotropic diffusion ($ D_x = D_y $) and anisotropic diffusion ($ D_x = 3D_y $) along the boundary.} \textbf{A}: Schematic of elliptical boundaries with varying aspect ratios and the corresponding empirical adatom density fields. \textbf{B--C}: Radial statistics extracted from ellipses with different aspect ratios under isotropic diffusion ($ D_x = D_y $) and anisotropic diffusion ($ D_x = 3D_y $), respectively. The semi-major axis $ a $ is oriented parallel ($ a > b $), at a $ 45^{\circ} $ tilt ($ a < b $), and orthogonal ($ a < b $) to one of the diffusion axes $ \mathbf{\hat{x}} $. By incorporating local curvature contributions, the resulting mean-field analytical solutions reproduce key features observed in the Monte Carlo–sampled empirical density fields. } 
    \label{fig4}
\end{figure*}

 \begin{figure*}[t!]
    \centering
    \includegraphics{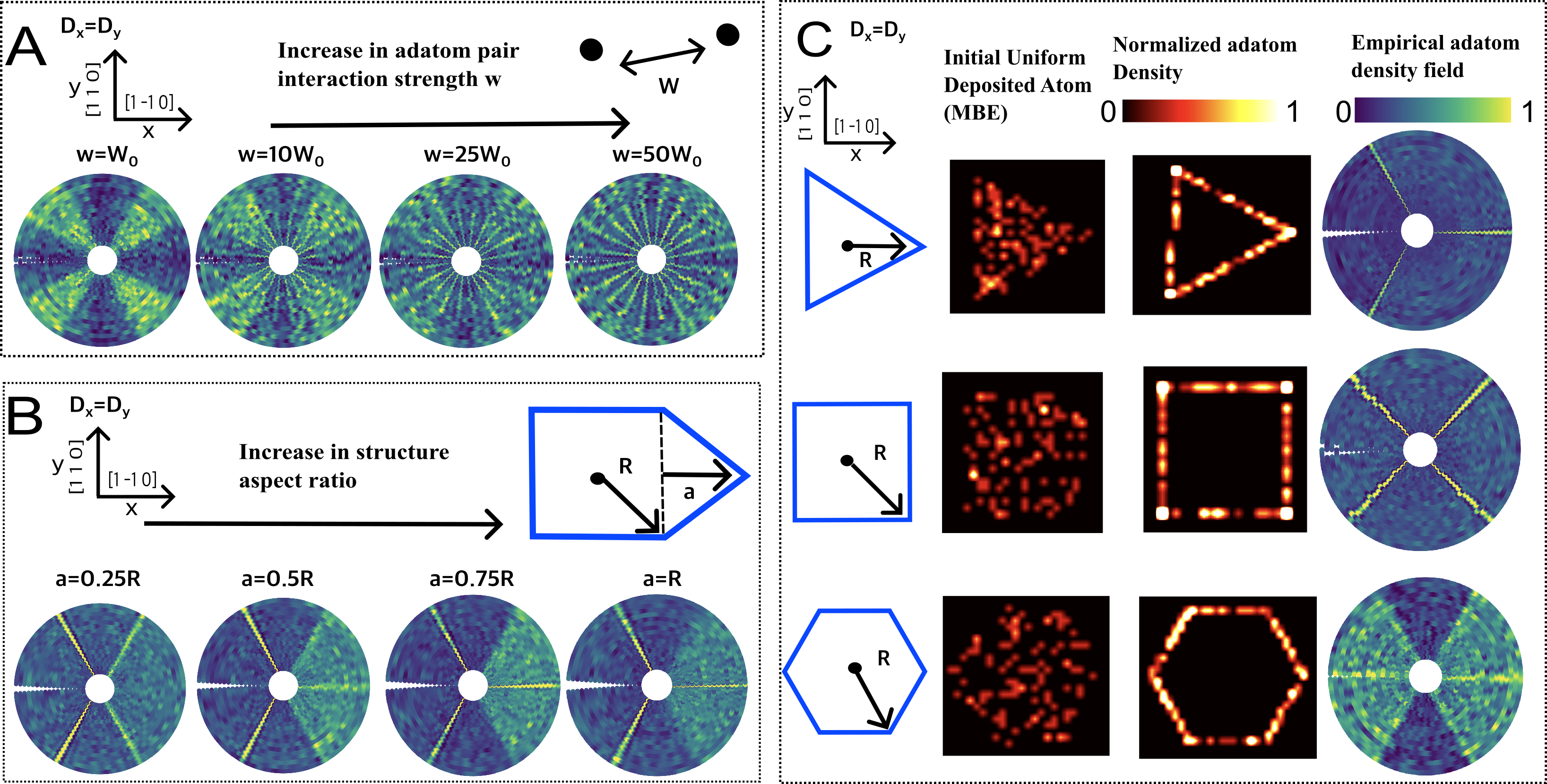}
    \caption{ {Geometric tuning of empirical adatom density fields under isotropic diffusion ($ D_x = D_y $).} \textbf{A}: By incorporating a nonzero repulsive pairwise interaction potential $ w $ between adatoms, the resulting empirical adatom density field exhibits a smeared-out asymptotic density profile along the boundary contour. \textbf{B}: Starting from a lozenge boundary, elongation of the right-side flat edge with an aspect ratio $ a/R $ induces local anisotropic confinement, leading to enhanced adatom accumulation near the elongated region. This demonstrates that local curvature can modulate the spatial profile of the adatom density field along the boundary. \textbf{C}: Simulations with various boundary geometries under initially uniform atomic deposition illustrate the flexibility and adaptability of the modeling framework in capturing geometry-tunable adatom density profiles and QD nucleation behavior along the boundary contour.} 
    \label{fig5}
\end{figure*}

 \begin{figure*}[t!]
    \centering
    \includegraphics{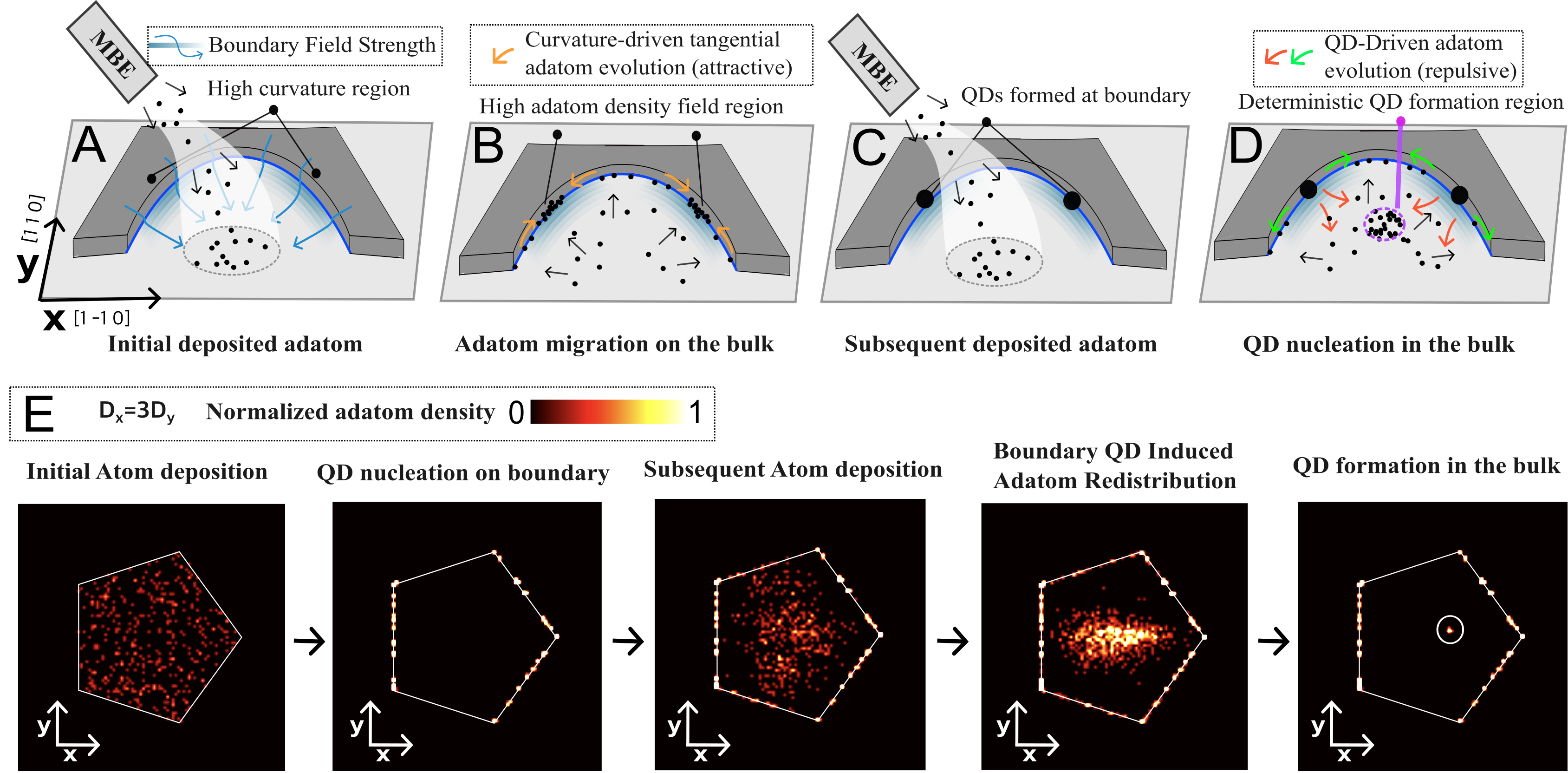}
    \caption{{Schematic of semi-deterministic QD placement guided by a boundary-induced chemical potential landscape.} \textbf{A}: Atoms are vertically deposited via an epitaxy source onto a patterned substrate with an embedded boundary field (field strength illustrated by the blue gradient), as defined in Fig.~\ref{fig1}. \textbf{B}: Anisotropic diffusion drives adatoms toward the boundary, where their motion becomes confined upon arrival. As described in Eq.~\ref{pde}, adatoms preferentially cluster in regions of high positive curvature (e.g., concave segments with compressive strain), increasing the local empirical adatom density (black dots) and promoting QD nucleation at these locations. \textbf{C}: Following initial QD formation at the boundary, additional atoms are vertically deposited via continued epitaxy at a much lower rate than the rate at which adatoms reach steady state. \textbf{D}: The presence of boundary QDs reshapes the local chemical potential landscape. Repulsive interactions from existing QDs redirect adatom migration either tangentially along the boundary (green arrows) or outward into the substrate interior (red arrows). These combined effects promote secondary QD nucleation within the bulk, enabling semi-deterministic and spatially reproducible QD patterning mediated by boundary-induced adatom dynamics. \textbf{E}: Snapshots from Monte Carlo simulations where the pentagon boundary profile is used to illustrate each process step described above. Circles indicate stable QDs, while red dots represent adatoms. } 
    \label{fig6}
\end{figure*}

 \begin{figure*}[t!]
    \centering
    \includegraphics{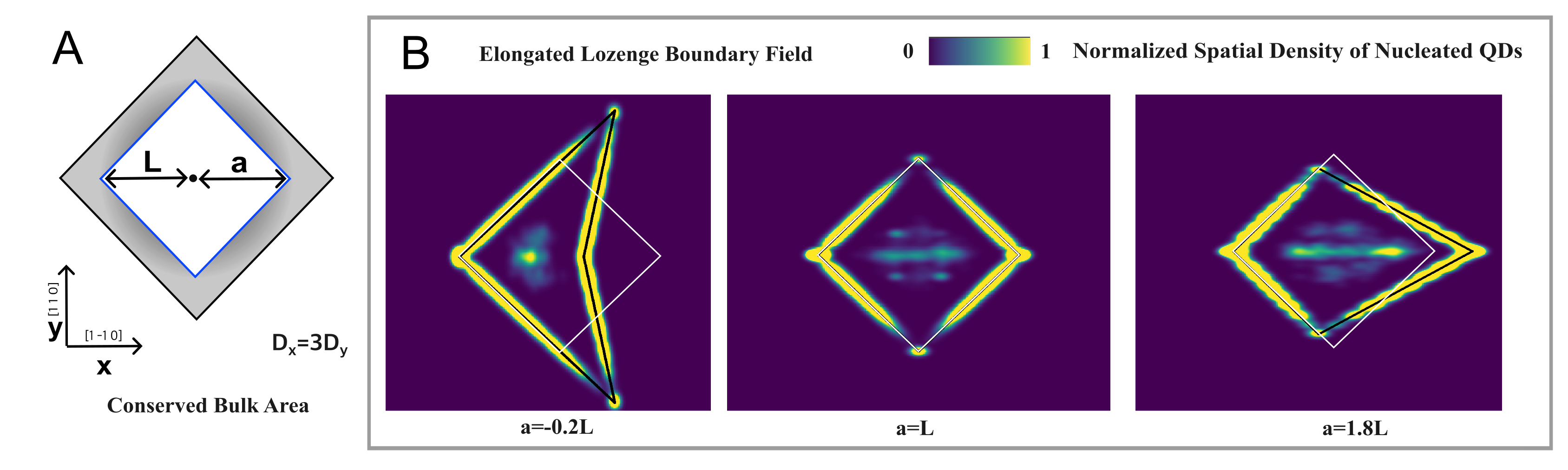}
    \caption{{Boundary-mediated semi-deterministic QD placement via full-scale Langevin dynamics ($ D_x = 3\,D_y $).} \textbf{A}: A lozenge domain with a tunable right-side diagonal coordinate $ (a, 0) $, where $ a \in (-0.6L, 2L) $, is used while keeping the total bulk area constant. \textbf{B}: Spatial distribution of the first QD nucleation site within the bulk for different aspect ratios, sampled from statistics over 400 distinct random adatom initializations (i.e. independent with different initial conditions). Although QDs nucleate stochastically, the resulting distribution exhibits statistically localized behavior that is tunable solely through boundary field design, demonstrating the principle of semi-deterministic QD placement. A detailed parameter sweep is provided in the Appendix D, Fig.~\ref{fig8}. We note that the above 'QD distributions' are independently sampled statistics based on different initial adatom density profiles (i.e., they are not clusters of QDs), and they represent realizations that reflect the probability of possible QD nucleation sites. } 
    \label{fig7}
\end{figure*}

{In Figs.~\ref{fig3}, \ref{fig4}, and \ref{fig5}, using Eq.~\ref{nc}, we systematically investigate QD nucleation profiles in the diluted limit across various boundary geometries, where stochastic seeding events occur almost exclusively at the boundary. The seeding probability is modulated by the adatom density field, which reflects the combined influences of biaxial diffusion, boundary field effects, and curvature. In the simulations, the ensemble-averaged empirical adatom density field $ \langle \hat{n} \rangle $ is obtained from Langevin-based Monte Carlo sampling.}

As shown in Fig.~\ref{fig3}, this comparison is performed across multiple concentric circular boundaries of varying radii, under both isotropic and anisotropic diffusion conditions (Fig.~\ref{fig3}B and \ref{fig3}C). When the QD nucleation potential $ V_{\mathrm{nc}} $ is included, QDs are distributed along the boundary with spatial profiles that closely resemble the empirical adatom density fields. The spacing between boundary QDs is determined by the interplay of diffusion, geometry, and $ V_{\mathrm{nc}} $. For instance, once a QD nucleates at a given location along the boundary, it increases the local chemical potential, generating effective repulsive interactions that redirect subsequent adatoms to other regions.

In Fig.~\ref{fig3}D, we construct the analytical solutions of the coarse-grained model under the mean-field approximation and compare them with empirical adatom density results, demonstrating quantitative agreement in capturing the stationary spatial features.

In both isotropic and anisotropic cases, peaks in the adatom density field correspond to regions of enhanced seeding probability, where adatoms are more likely to form temporary clusters. These clusters represent the initial stage of QD formation. Once temporary clusters form in regions of high density, the seeding potential $ V_{\mathrm{seed}} $ attracts additional adatoms deposited during heteroepitaxy, leading to the formation of stable QDs along the boundary.

To further quantify the contribution of curvature to the adatom density field along the boundary beyond the case of uniform curvature, we extend the above formulation to elliptical boundary geometries (Appendix C) and shown in Fig.~\ref{fig4}A. In Fig.~\ref{fig4}B,C, we construct analytical solutions of the coarse-grained model under the mean-field approximation for different axial orientations under both isotropic and anisotropic diffusion, and compare them with the corresponding empirical density fields obtained from Langevin-based Monte Carlo simulations. In regions where the empirical adatom density becomes locally elevated, QD seeding events become statistically favorable. This effect, combined with the fact that regions of high curvature act as effective traps for adatoms, suggests that both biaxial diffusion and local curvature can serve as tuning parameters to control adatom density fields and thus QD nucleation sites along boundary contours. Since boundary geometries can be arbitrarily fabricated, as illustrated in Fig.~\ref{fig5}, this approach enables flexible design strategies for device applications.

As the total number of deposited atoms increases, repulsive pairwise interactions $ w $ (see Eq.~\ref{pair eq}) become non-negligible. This results in a modification of the steady-state solution, $ n_{\mathrm{int}}(x,y) \sim n(x,y)\exp[q(w)] $, where $ q(w) $ is a functional of the interaction strength that broadens the density field along the boundary contour. Intuitively, as repulsive interactions increase, adatoms tend to avoid regions of high density, resulting in a smeared out density profile along the boundary, as shown in Fig.~\ref{fig5}A. The governing PDE in Eq.~\ref{pde} thus becomes coupled through the pairwise interaction term, and nontrivial spatial correlations arising from local curvature and boundary nucleation potentials must be taken into account. 

\zw{As additional adatoms are deposited, they become ``aware'' of the surrounding strain field, generated by both the engineered boundary and primary QDs nucleated at high-curvature boundary regions. This evolving strain landscape dynamically alters subsequent QD nucleation sites at the boundaries and within the bulk (pristine surface). Building on this, in the following section we utilize strain sensitivity, together with engineered boundary fields, to guide QD nucleation in a controlled manner, enabling semi-deterministic QD placement.}

\section{\rom{5}. secondary QD formation on pristine surface}
A stable InAs/GaAs QD on a pristine surface has a typical size range of 20 to 25~nm with nucleation sites that are both highly sensitive to material properties and growth conditions. The interaction length scale of the nucleation potential $V_{\mathrm{nc}}$ can extend to several tens of nanometers, or even up to hundreds of nanometers, as inferred from the observed relationship between QD density and growth rate, or QD array periodicity  ~\cite{Konishi2017,mano2002formation}. In contrast, stable QDs formed at the boundary often exhibit irregular shapes and accumulated stress profiles, which can result in comparable or even greater interaction length scales, ranging from a few nanometers to several hundred nanometers. The repulsive field generated by those stable QDs at the boundary not only affects the migration of nearby adatoms but also interacts with adatom densities in the bulk (pristine surface), influencing the global spatial accumulation and redistribution of newly deposited adatoms, which can be used to statistically control the seeding location on the pristine bulk surface. This concept is illustrated schematically in Fig.~\ref{fig6}.

As demonstrated in Fig.~\ref{fig7}A, to model boundary-mediated secondary QD nucleation, we begin with a elongated lozenge boundary field design and vary the right-side diagonal coordinate $ (a, 0) $ from $ -0.6L $ to $ 2L $, while conserving the total bulk area. The epitaxial deposition of new atoms is introduced sequentially at a rate much slower than the time required for adatoms to reach a steady-state distribution through Langevin dynamics. We then record the spatial distribution corresponding to the \textit{first} QD nucleation site, as illustrated on the far right side of Fig.~\ref{fig6}E. This process is repeated 400 times using different initial random adatom configurations (uniformly distributed). Depending on the value of $ a $, as shown in Fig.~\ref{fig7}B and Fig.~\ref{fig8}, the resulting spatial distribution of stable QDs exhibits a boundary-dependent profile. The observed inhomogeneity in the QD distribution reveals statistically preferential nucleation sites. As detailed in Appendix D Fig.~\ref{fig9}, we further perform a full parameter sweep over $ a \in [-0.6L, 2L] $ for a hexagonal boundary field with a $ 60^\mathrm{o} $ orientation. While the microscopic parameters of the Langevin dynamics remain unchanged, the hexagonal boundary field produces a more spatially extended adatom density field compared to the elongated lozenge boundary field shown in Fig.~\ref{fig8}.

\zw{Intuitively, semi-deterministic placement is expected: upon achieving primary QD distribution along the engineered boundary field, subsequent adatoms mainly diffuse along the fast axis (and hardly along the slow axis) and experience secondary interactions that can be significantly slowed down when near primary QDs, leading to adatom accumulation over time at those locations. }

\zw{While the above simulation is done at mild anisotropy, $D_x = 3D_y$, for illustration purposes, in the case of the GaAs(001)-$\beta_2 (2\times 4)$ surface with diffusion axes $\mathbf{\hat{x}} \equiv [1\bar{1}0]$ and $\mathbf{\hat{y}} \equiv [110]$, the diffusion constant and anisotropy, $D_x \sim 10^3 D_y \sim 1 \times 10^{-3}~\mu\mathrm{m}^2/\mathrm{s}$ along the two directions, follow at typical temperatures $T \sim 500~\mathrm{K}$ \cite{rosini2009indium}, which gives the relative diffusion length scale  $\ell_x/\ell_y\sim 32$. This suggests that an efficient redistribution of the empirical density field occurs along the fast diffusion axis. Therefore, we predict that by engineering boundary fields with anisotropy along the fast diffusion axis (for example, shape edges across two opposite boundaries, see the comparison in Appendix D, Fig.~\ref{fig8} and Fig.~\ref{fig9}), the modulation effects from both the boundary fields and the on-boundary primary nucleated QDs would exert the strongest influence on subsequently deposited adatoms, leading to more pronounced modulation that enables semi-deterministic placement of secondary QDs. }

\zw{Since we assume a quasi-stationary limit, the deposition rate must be sufficiently low to allow adatoms to reach quasi-equilibrium. Let's consider a unit deposition time, with adatom diffusion lengths of $\ell_x \sim 32 \,\mathrm{nm}$ and $\ell_y \sim 1 \,\mathrm{nm}$. The effective diffusion area is then $d\Omega \sim \ell_x \cdot \ell_y = 32 \,\mathrm{nm}^2$. The deposition flux $F$ (atoms per unit area per unit time) corresponding to $N$ atoms deposited in this area is given by $F \sim N/d\Omega$. In the quasi-stationary limit, within a unit time, the number of deposited adatoms over the effective area should be much smaller than one adatom per diffusion area. Taking the bound $N=1$, the corresponding flux is $F = 1/(32\,\mathrm{nm}^2 \cdot 1\,\mathrm{s}) \approx 3.125 \times 10^4\,\mathrm{atoms/(\mu m^2 \cdot s)}$. To convert this to a deposition rate in monolayers per second (ML/s), we use the lattice constant of GaAs ($a=0.565\, \mathrm{nm}$) to calculate the surface atomic density as $n_s = 2/a^2 \sim 6.26 \times 10^6\,\mathrm{atoms/\mu m^2}$. This gives the upper bound on the deposition rate $R_\mathrm{bound} = F/n_s \sim 5\times 10^{-3}\,\mathrm{ML/s}$, which is consistent with typical growth rates used for InAs QDs on GaAs(001) substrates \cite{Vullum2017}. To achieve the quasi-stationary limit, the empirical deposition rate must satisfy $R \ll R_\mathrm{bound}$ around temperature $T\sim 500$ K, which can be realized under experimental conditions.}

\zw{Practically, designing such boundary fields is straightforward with standard fabrication capabilities. A typical hard mask can first be uniformly deposited across the substrate (e.g., a (001)-oriented GaAs surface), and then selectively etched to re-expose patterned regions for regrowth.}

 \section{\rom{6}. Discussion and Conclusion}
 
\zw{Over the past few decades, strain-mediated interactions between QDs and adatoms have been extensively studied theoretically, focusing on QD morphology and phase transitions, yet few studies have explored their use in practical applications such as QD placements.} In this work, we have theoretically and numerically investigated the interplay between geometry, diffusion, and stochastic dynamics in the context of quantum dot (QD) nucleation from empirical adatom density fields. Our results illustrate how geometric properties of boundary contours, such as curvature and orientation relative to anisotropic diffusion axes, along with intrinsic microscopic properties, such as adatom interactions and QD nucleation dynamics, shape the spatial patterns of QD formation. 

In the early stage of epitaxy, referenced as the diluted limit, adatoms are preferentially accumulated along the boundary contour where curvature enhances the local adatom density fields, which can be described by the path integral formulation of a coarse-grained theory. As nucleated QDs accumulate on the boundary, their presence introduces a repulsive interaction field that dynamically alters the chemical potential landscape within the bulk. This feedback mechanism effectively shifts the nucleation zone away from the initial boundary, giving rise to secondary QD populations within the bulk, as demonstrated numerically via the fully scale Langevin dynamics simulations.

While this effect has not yet been experimentally realized, it has broad implications for nanofabrication strategies where precise spatial control of QDs is required in conjunction with complex device architectures. For example, on-chip indistinguishable multi-photon sources require simultaneous electrical tuning and coordination to dynamically bias individual QDs into their trion states, which is essential for constructing large-scale photonic cluster states, a key component of measurement-based quantum computing~\cite{PhysRevLett.116.020401,Zhai2020,Cogan2023,Huet2025}. Furthermore, the controlled spatial arrangement of QDs enables systematic investigations of decoherence, dissipation, and entanglement propagation in open quantum systems~\cite{PhysRevA.110.032407,MOHAMED2019125905}, where spatial correlations and engineered environments play a central role~\cite{Huang2025,PhysRevA.57.120}.

This work enables the semi-deterministic positioning of high optical quality self-assembled quantum dots on pristine surfaces, opening a new direction for next-generation on-chip photonic devices.

\section{Acknowledgement}
We acknowledge support from NSF Award No. 2427169, 2137740 and Q-AMASE-i, through Grant No. DMR-1906325, and from NWO Quantum Software Consortium (Grant No. 024.003.037). Furthermore, the authors thank Chen Shang, Paul J. Simmonds, and John E. Bowers for their insightful discussions and suggestions.

\medskip
\bibliographystyle{unsrt} 

\widetext

\appendix 
\section{Appendix A: Steady-state Solutions }

We first assume $\kappa=0$, and as an example, the curvature-free steady-state solution in x-direction is given by, 
\begin{equation}
T_x^2 D_x \frac{\partial^2 f}{\partial x^2} - \lambda f = 0, \quad f(x) \sim e^{k_x^{\pm} |x-x_0|}, \qquad T_x^2 D_x\, (k_x^{\pm})^2 - \lambda = 0 \quad \Longrightarrow \quad (k_x^{\pm})^2 = \frac{\lambda}{T_x^2 D_x}.
\end{equation}
In general, the coefficients presented in Eq.~\ref{steady-state} have the following form, 
\begin{equation}
\begin{aligned}
{k}^{\pm}_v=   &    \frac{{\kappa T_v D_v} \pm \sqrt{{(\kappa T_v D_v) }^2 + 4T^2_v D_v \lambda}}{2T^2_v D_v}.\\
\end{aligned}
\end{equation}

\section{Appendix B: Path Integral Formulation and Mean-field Approximation }

Starting from a reference point on the boundary, we can obtain a general solution for the coarse-grained adatom density field by considering the following,
\begin{equation}
\begin{aligned}
\frac{n(\theta_{N}) }{n_0  } &= \prod_{i=0}^{N} \sum_{c,c'=\pm 1} A^i_{c,c'} \exp\left( k_x^{c} |x(\theta_{i+1}) - x(\theta_{i})| + k_y^{c'} |y(\theta_{i+1}) - y(\theta_{i})| \right) \\
&= \prod_{i=0}^{N} \sum_{c,c'=\pm 1} A^i_{c,c'} \exp\left( du_i(\, k_x^{c} |T_x(u_i)| + k_y^{c'} |T_y(u_i)|) \right).
\end{aligned}
\end{equation}

We further simplify the above by noting that,
\begin{equation}
   k_x^{c} |T_x(u_i)| = \frac{\kappa}{2} + c\, \sqrt{\left(\frac{\kappa}{4} \right)^2 + \ell_x^{-2}}, \qquad 
   k_y^{c'} |T_y(u_i)| = \frac{\kappa}{2} + c'\, \sqrt{\left(\frac{\kappa}{4} \right)^2 + \ell_y^{-2}},
\end{equation}
where $ \ell_x^{-1} = \sqrt{\lambda/D_x} $ and $ \ell_y^{-1} = \sqrt{\lambda/D_y} $ represent the diffusion lengths along the $ x $- and $ y $-directions, respectively.

Substituting these expressions, we obtain,
\begin{equation}
\begin{aligned}
\frac{n(\theta_{N}) }{n_0  } &= \prod_{i=0}^{N} \sum_{c_i,c_i'=\pm 1} A^i_{c_i,c_i'} \exp\left[ du_i \left( \kappa(u_i) + c_i\, \sqrt{\left(\frac{\kappa(u_i)}{4} \right)^2 + \ell_x^{-2}} + c_i'\, \sqrt{\left(\frac{\kappa(u_i)}{4} \right)^2 + \ell_y^{-2}} \right) \right] \\
&= \sum_{\mathbf{c},\mathbf{c'}}  A_{\mathbf{c},\mathbf{c'}} \prod_{i=0}^{N} \exp\left[ du_i \left( \kappa(u_i) + \mathbf{c}_i\, \sqrt{\left(\frac{\kappa(u_i)}{4} \right)^2 + \ell_x^{-2}} + \mathbf{c}'_i\, \sqrt{\left(\frac{\kappa(u_i)}{4} \right)^2 + \ell_y^{-2}} \right) \right],
\end{aligned}
\end{equation}
where, $    \mathbf{c} = [1, -1, 1, 1, \dots], \quad \mathbf{c'} = [-1, -1, 1, -1, \dots]$, are sequences (or path) of signs $ c_i $ indexed by $ i = 0, \dots, N $. The summation is taken over all possible combinations of $ \mathbf{c} $ and $ \mathbf{c'} $, representing distinct sign trajectories (path) within the exponential.

In the continuum limit, the product becomes an integral in the exponent, leading to the following path integral representation,
\begin{equation}
\begin{aligned}
\frac{n(u) }{n_0  } = \int \mathcal{D}[\mathbf{c}] \, \mathcal{D}[\mathbf{c'}] \, A[\mathbf{c}, \mathbf{c'}] \exp\left[ \int_{0}^{u} d\tilde{u} \left( \kappa(\tilde{u}) + \mathbf{c}(\tilde{u})\, \mathcal{K}_x(\tilde{u}) + \mathbf{c'}(\tilde{u})\, \mathcal{K}_y(\tilde{u}) \right) \right],
\end{aligned}
\end{equation}
as discussed in Eq.~\ref{path-integral}.

The above path integral formulation can be numerically approximated under a mean-field assumption by directly evaluating the defining expression,
\begin{equation}
\begin{aligned}
n(\phi,\theta_0)  &\sim \sum_{\mathbf{c},\mathbf{c'}} \exp\left[ \sum_{i=0}^{\phi} \kappa(u_i) \left( x(\theta_{i+1}) - x(\theta_{i}) + y(\theta_{i+1}) - y(\theta_{i}) \right) \right] \\
&\approx \exp\left[ \kappa(\phi) \left( R(D_x)\big(x(\phi) - x(\theta_0)\big) + R'(D_y)\big(y(\phi) - y(\theta_0)\big) \right) \right] \\
&\quad + \exp\left[ -\kappa(\phi) \left( R(D_x)\big(x(\phi) - x(\theta_0)\big) + R'(D_y)\big(y(\phi) - y(\theta_0)\big) \right) \right],
\end{aligned}
\label{mean field}
\end{equation}
where $ R(D_x) $ and $ R'(D_y) $ are signed (parametrization-dependent) coefficients. In the last line, only two overall sign contributions are retained, consistent with the endpoint (mean-field) treatment of the path sum.

To obtain the ensemble-averaged density, we further sum over all possible starting points $ \theta_j $, 
\begin{equation}
    n(\phi) \sim \sum_{j=0}^{2\pi} n(\phi, \theta_j).
\end{equation}
The above mean-field solutions are numerically demonstrated on circular and elliptical boundary fields presented in Fig.~\ref{fig3} and Fig.~\ref{fig4}.

\section{Appendix C: Elliptical boundary parametrization }

Consider an isotropic diffusion setting with $ D_x = D_y $, and an elliptical boundary confinement centered at the origin, characterized by a semi-major axis $ a $ and a semi-minor axis $ b $ (as shown in Fig.~\ref{fig4}A). The boundary is parametrized by $ \theta_0 \in [0, 2\pi) $, corresponding to a boundary point located at $ [a\cos\theta_0,\, b\sin\theta_0] $. 

For simplicity, we define the local configuration space using a second ellipse with the same aspect ratio, scaled by a global constant $ r $, such that,
\begin{equation}
    \mathbf{r}_{\mathrm{bd}}(r,\phi,\theta_0) = a(r\cos\phi - \cos\theta_0)\,\mathbf{\hat{x}} + b(r\sin\phi - \sin\theta_0)\,\mathbf{\hat{y}}.
\end{equation}

The local curvature at the boundary point (i.e., when $ r = 1 $) takes the form,
\begin{equation}
    \kappa(\theta_0) = \frac{ab}{\left( b^2 \cos^2\theta_0 + a^2 \sin^2\theta_0 \right)^{3/2}}. 
\end{equation}

Analytical solutions of the adatom density field under the mean-field approximation (Eq.~\ref{mean field}) show close agreement with the results obtained from direct simulations of adatom Langevin dynamics sampled via Langivn-based Monte-Carlo (Fig.~\ref{fig4}B,C).

\clearpage
\section{Appendix D: Boundary-mediated semi-deterministic QD placements under different geometries }

\begin{figure*}[h!]
    \centering
    \includegraphics{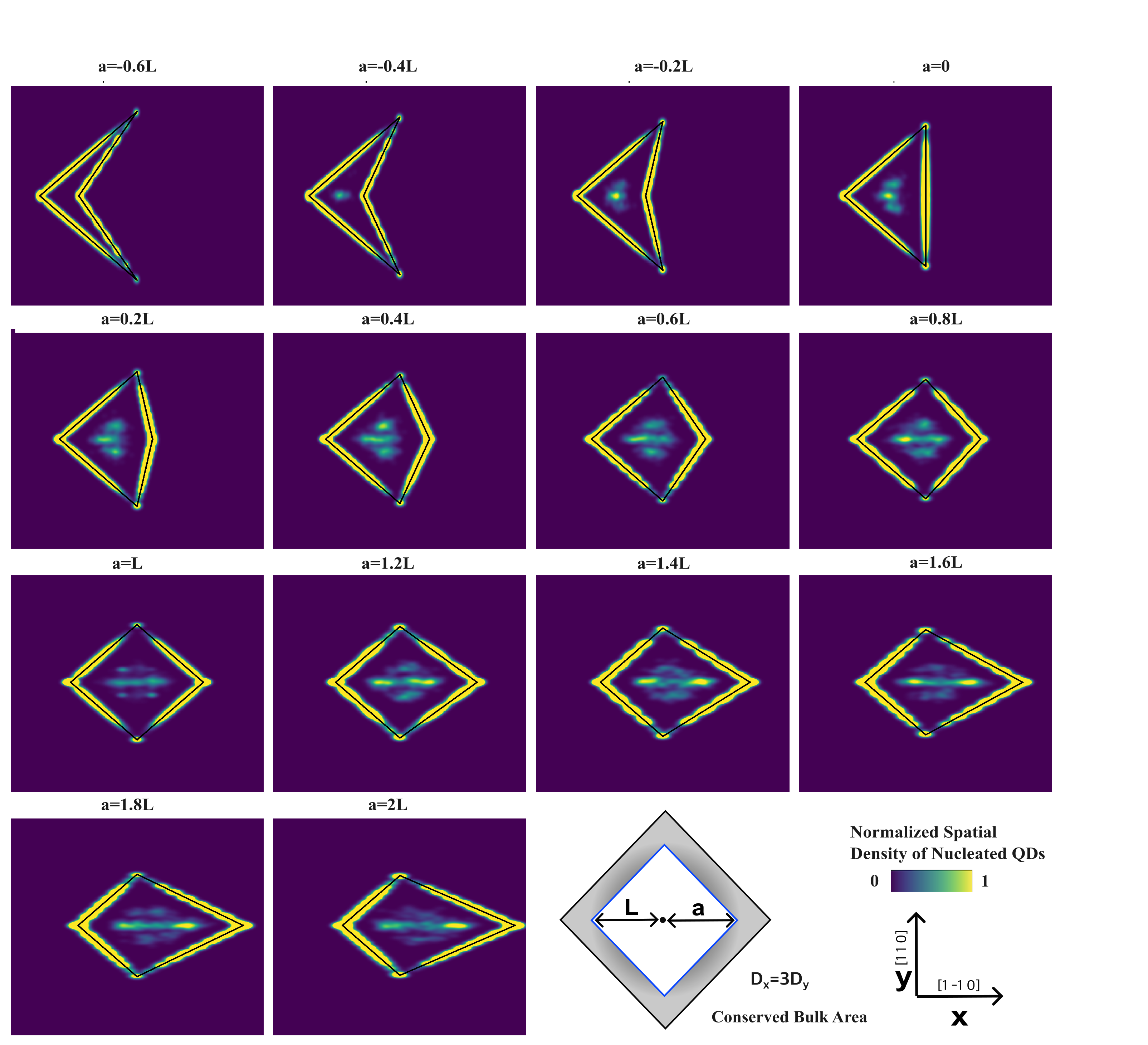}
    \caption{ Full parameter sweep over $a\in [-0.6L,2L]$ for the elongated lozenge boundary field presented in Fig.~\ref{fig7} }.  
    \label{fig8}
\end{figure*}

\begin{figure*}[h!]
    \centering
    \includegraphics{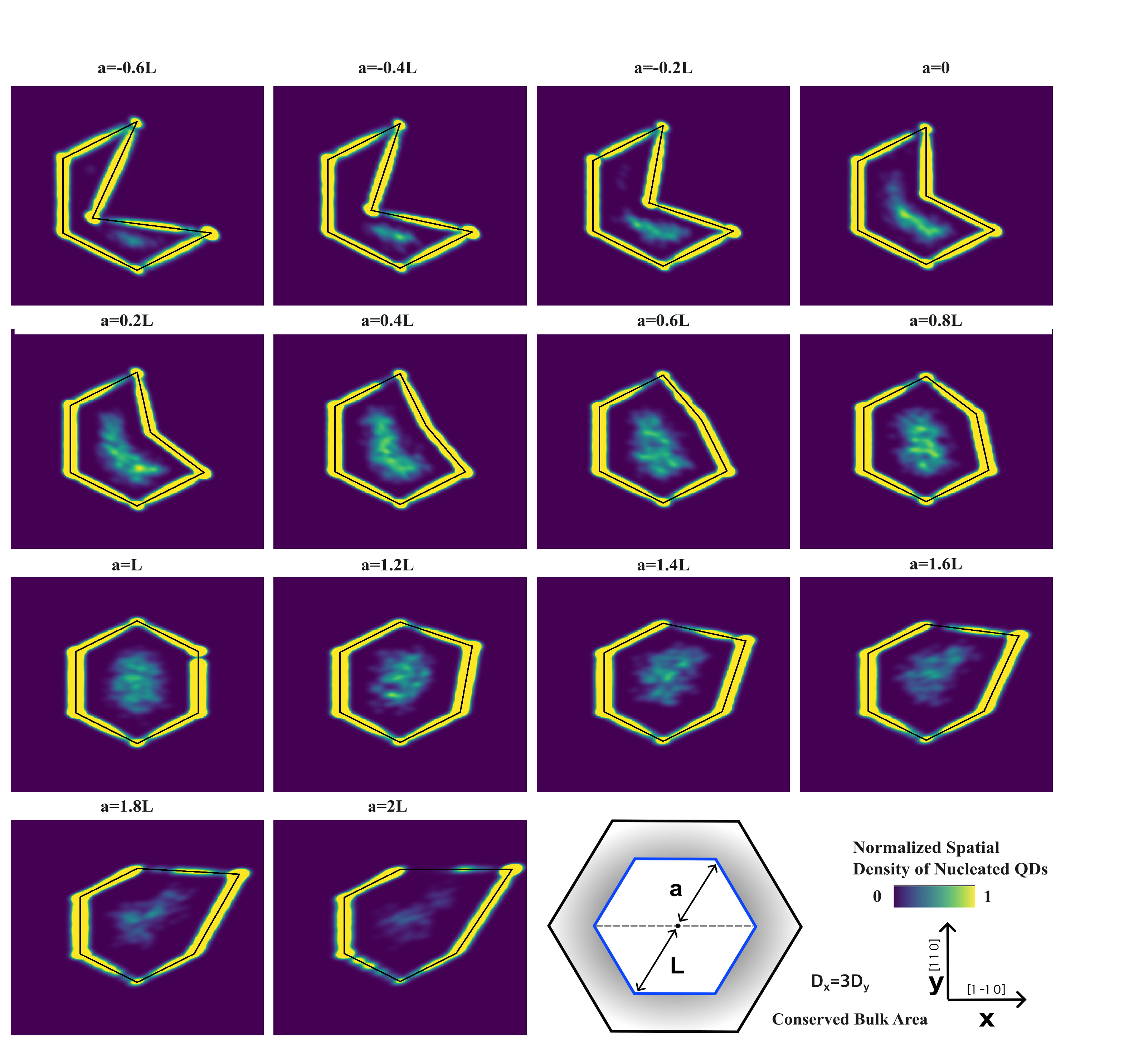}
    \caption{ A full parameter sweep over $a\in [-0.6L,2L]$ for the hexagonal boundary field at $60^\mathrm{o}$ angle.}  
    \label{fig9}
\end{figure*}

As an extension of Fig.~\ref{fig7}, we illustrate how variations in the elongated lozenge boundary field (aspect ratio $a$) affect the spatial QD distribution along the boundary contour and secondary nucleation statistics (Fig.~\ref{fig8}). Negative $a$ values produce compressed nucleation statistics with enhanced localization near the lozenge vertices along the fast diffusion axis, and increasing $a$ elongates the secondary QD nucleation distribution. In contrast, due to flat edges, the hexagonal boundary field at a ${60}^\mathrm{o}$ angle (Fig.~\ref{fig9}) exhibits weaker statistical localization than the lozenge geometries.

\end{document}